\documentclass[11pt]{article}
\usepackage{amsmath, amssymb, graphics}

\usepackage[nosort]{cite}

\usepackage{amsbsy}
\usepackage{amsfonts}
\usepackage{amsmath}
\usepackage{graphicx}
\usepackage{array}
\usepackage{booktabs}
\usepackage{float}
\usepackage[usenames, dvipsnames]{color}
\usepackage{comment}
\usepackage{hyperref}
\numberwithin{equation}{section}
\usepackage{slashed}

\usepackage{appendix}
\usepackage[nodayofweek]{date time}

\newcommand{\mathsym}[1]{{}}

\def\be{\begin{eqnarray}}
\def\ee{\end{eqnarray}}

\makeatletter
\renewcommand\section{\@startsection {section}{1}{\z@}%
                                   {-3.5ex \@plus -1ex \@minus -.2ex}%
                                   {2.3ex \@plus.2ex}%
                                   {\normalfont\large\bfseries}}
\renewcommand\subsection{\@startsection{subsection}{2}{\z@}%
                                   {-3.25ex\@plus -1ex \@minus -.2ex}%
                                   {1.5ex \@plus .2ex}%
                                   {\normalfont\normalsize\bfseries}}

\makeatother

\def \tr {{\rm tr}}

\def\det{\hbox{det}}

\def \varpi {{\rm w}}

\newcommand{\preprint}[1]{\begin{table}[t]  
             \begin{flushright}               
             {#1}                             
             \end{flushright}                 
             \end{table}}                     
\renewcommand{\title}[1]{\vbox{\center\LARGE{#1}}\vspace{5mm}}
\renewcommand{\author}[1]{\vbox{\center#1}\vspace{5mm}}
\newcommand{\address}[1]{\vbox{\center\em#1}}

\begin{document}
\begin{titlepage}
\date{\today}

\preprint{PUPT-2530}

\begin{center}
\hfill \\
\hfill \\
\hfill \\
\hfill \\

\title{Testing the Boson/Fermion Duality on the Three-Sphere}

\author{Simone Giombi}

\address{Department of Physics, Princeton University, Princeton, NJ 08544, USA}

\end{center}

\vskip 1cm

\abstract{We study the duality between theories of a fundamental scalar or fermion coupled to $U(N)$ Chern-Simons gauge theory at the level 
of the three-sphere partition function, or equivalently entanglement entropy across a circle. 
The duality relation between the sphere free energies of the large $N$ bosonic and fermionic conformal 
theories is sensitive to certain shifts in the Chern-Simons level already at order $N$. We first study 
a similar duality in a ${\cal N}=2$ supersymmetric Chern-Simons matter theory, where it can be checked exactly using localization. 
At large $N$, the free energies of supersymmetric and non-supersymmetric theories are related in a simple way, and we use this fact to 
obtain an explicit solution for the scalar and fermion free energies which obey the duality map and have the expected weak and strong coupling behaviors. 
We also suggest that a worldline representation of the free energy allows to relate its calculation to a sum of Wilson loops in pure Chern-Simons 
theory. In the supersymmetric case, we find that this approach precisely reproduces the localization prediction. 
In the non-supersymmetric case, we observe that a result consistent with the boson/fermion duality can be obtained using a certain framing prescription, which remains to be understood. We also briefly discuss the case of massive theories, and compare the small mass expansion of the free energy to previously known 
results for the correlation functions of bilinear operators in flat space.}

\vfill

\end{titlepage}

\eject \tableofcontents

\setcounter{footnote}{0}

\section{Introduction and Summary}

There is growing evidence that the three-dimensional conformal field theories obtained by coupling Chern-Simons (CS) gauge theory to 
scalars or fermions in the fundamental representation obey a non-perturbative duality relating the bosonic and fermionic theories. Essentially, 
this amounts to a version of bosonization for non-abelian conformal gauge theories in 3d. First suggestions of this boson/fermion duality were made in 
the context of studying the large $N$ limit of these theories and their AdS/CFT duals \cite{Giombi:2011kc, Aharony:2011jz, Chang:2012kt}. In particular, 
it was proposed that the bosonic and fermionic vector models are dual to parity breaking versions of Vasiliev higher-spin theory \cite{Vasiliev:1992av},
generalizing the earlier conjectures of \cite{Klebanov:2002ja,Leigh:2003gk,Sezgin:2003pt} for bosonic and fermionic vector models without CS interactions. The 
boson/fermion duality is suggested by the structure of the dual higher-spin theories,\footnote{See e.g. \cite{Chang:2012kt, Giombi:2012ms, Giombi:2016ejx}
for reviews of higher spin/vector model dualities. Explicit calculations of some 3-point functions in the parity breaking HS theories that match 
the expected results in CS-vector models were given in \cite{ Giombi:2012ms}. Recently, further systematic tests were performed in \cite{Vasiliev:2016xui, Sezgin:2017jgm, Didenko:2017lsn}, following the field redefinition proposal made in \cite{Vasiliev:2016xui}, see also \cite{Gelfond:2017wrh, Vasiliev:2017cae}.} and also by the 
constraints imposed by the weakly-broken higher-spin symmetries at large $N$ \cite{Maldacena:2011jn, Maldacena:2012sf}. 
A first concrete suggestion for the duality map at finite $N$ 
was made in \cite{Aharony:2012nh}. 
Studying the mapping of monopoles and baryons \cite{Radicevic:2015yla}, as well as consistency with known level/rank dualities 
in pure CS theory, it was then refined and generalized in \cite{Aharony:2015mjs, Hsin:2016blu}, resulting in the following finite $N$ duality maps involving 
CS theory with unitary gauge group: 
\begin{eqnarray}
\!\!\!\!\!\!\!\!\!\mbox{$U(N)_{k,k}$ coupled to $N_f$ scalars}~~~&\leftrightarrow&~~~\mbox{$SU(k)_{-N+\frac{N_f}{2}}$ coupled to $N_f$ fermions}
\label{d1}\\
\!\!\!\!\!\!\!\!\!\mbox{$SU(N)_{k}$ coupled to $N_f$ scalars}~~~&\leftrightarrow&~~~\mbox{$U(k)_{-N+\frac{N_f}{2},-N+\frac{N_f}{2}}$ coupled to $N_f$ fermions}
\label{d2}\\
\!\!\!\!\!\!\!\!\!\mbox{$U(N)_{k,k\pm N}$ coupled to $N_f$ scalars}~~~&\leftrightarrow&~~~
\mbox{$U(k)_{-N+\frac{N_f}{2},-N\mp k+\frac{N_f}{2}}$ coupled to $N_f$ fermions}
\label{d3}
\end{eqnarray}
where the notation $U(N)_{k,k'}$ indicates that $k$ is the level of $SU(N)$ and $k'$ the level of the $U(1)$ factor.\footnote{More 
precisely, $U(N)_{k,k'} = (SU(N)_k \times U(1)_{Nk'})/Z_N$.} 
The integer $N_f$ is the number of matter fields in the fundamental representation of the gauge group, and the dualities are 
conjectured to hold for $N_f \le N$ (see \cite{Komargodski:2017keh} for a recent proposal on the behavior of these theories for $N_f > N$). 
It is implicit that the scalar theories include quartic self-interactions, and the duality applies to the corresponding IR fixed points (masses 
are tuned to zero to reach the conformal phase). 
The above dualities were recently also generalized in \cite{Aharony:2016jvv} to orthogonal and symplectic gauge groups, but we will not discuss 
these cases below. Note that for $N_f=0$ (or giving masses to the matter fields and flowing to the IR), 
all these boson/fermion relations reduce to the level/rank dualities of 
pure CS theory \cite{Naculich:1990pa, Mlawer:1990uv, Camperi:1990dk, Hsin:2016blu}, which are well established for finite $N, k$. 
For low $N$, 
the above boson/fermion dualities may find applications in condensed matter physics. In particular, in the abelian case  
(\ref{d1})-(\ref{d3}) were shown to be 
part of a web of dualities \cite{Seiberg:2016gmd, Karch:2016sxi,Murugan:2016zal} 
including the particle/vortex duality \cite{Peskin:1977kp, Dasgupta:1981zz}, as well as new fermion/fermion dualities 
\cite{Son:2015xqa,Wang:2015qmt, Metlitski:2015eka}. For further related work, see 
e.g. \cite{Radicevic:2016wqn,Kachru:2016rui,Kachru:2016aon, Filothodoros:2016txa, Karch:2016aux, Benini:2017dus,Nastase:2017gxr}. 

While there is evidence and several consistency checks that the dualities hold at finite $N$, 
it is hard to obtain explicit tests of them due to their non-perturbative nature. Many non-trivial calculations 
confirming the duality have nevertheless 
been obtained in the large $N$ 't~Hooft limit with $N/k$ fixed (and $N_f=1$ or held fixed in the large $N$ limit). 
These include matching correlation functions of 
local operators \cite{Aharony:2012nh,GurAri:2012is,Bedhotiya:2015uga, Giombi:2016zwa}, 
thermal free-energies \cite{Giombi:2011kc,Jain:2012qi,Aharony:2012ns,Jain:2013py,Takimi:2013zca} 
and $S$-matrices \cite{Jain:2014nza, Inbasekar:2015tsa, Yokoyama:2016sbx}. 
Further evidence also comes by relating  the 
non-supersymmetric dualities to well-established supersymmetric ones \cite{Giveon:2008zn, Benini:2011mf} via RG flow \cite{Jain:2013py,Gur-Ari:2015pca}. 
Being in the planar limit, all such tests do not distinguish between the different 
versions of the duality (\ref{d1})-(\ref{d3}), and also they are not sensitive to the half-integer shifts of the Chern-Simons level. 

In this note, we explore a different test of the duality, based on comparing the value of the sphere free-energy $F=-\log Z_{S^3}$ 
for dual theories, where $Z_{S^3}$ 
is the partition function on a Euclidean three-sphere with round metric. For a CFT, 
this is a well-defined physical quantity, which turns out to be related to the entanglement entropy 
across a circular entangling surface \cite{Casini:2011kv}. Its physical interest also stems 
from the fact that it provides an analog of the 2d $c$-theorem \cite{Zamolodchikov:1986gt} 
and 4d $a$-theorem \cite{Cardy:1988cwa, Komargodski:2011vj} that applies to 3d quantum field theories: 
if an RG flow trajectory connects a CFT in the UV to another CFT in the IR, then the sphere free energy satisfies $F_{UV}> F_{IR}$ 
\cite{Jafferis:2011zi, Klebanov:2011gs, Casini:2012ei} (see \cite{Pufu:2016zxm} and references therein for a recent review). 

In 
supersymmetric theories with ${\cal N}\ge 2$, the sphere partition function 
can be computed exactly by the method of supersymmetric localization 
\cite{Pestun:2007rz, Kapustin:2009kz, Jafferis:2010un}. This has proved to be an invaluable tool to obtain non-perturbative tests of various SUSY 
dualities \cite{Intriligator:1996ex, deBoer:1996mp, Aharony:1997gp, Giveon:2008zn,Benini:2011mf}, see \cite{Willett:2016adv} for a review and a more 
complete list of references. In non-supersymmetric interacting field theories, it is in general very hard to compute $F$ exactly. Here we will 
focus on the large $N$ `t~Hooft limit, where the existing calculations mentioned above provide evidence 
that the Chern-Simons vector models should be essentially exactly solvable at any value of the `t~Hooft coupling.\footnote{Large $N$ 
expansions of $F$ for other non-supersymmetric CFTs were developed in \cite{Klebanov:2011gs, Klebanov:2011td, Giombi:2015haa, Tarnopolsky:2016vvd}.}
We will focus on the case of $N_f=1$, but 
all of our calculations can be extended to higher $N_f$, as long as $N_f$ is held fixed in the large $N$ limit. As in previous large $N$ tests, our considerations 
in the planar limit do not distinguish between the different versions of the duality (\ref{d1})-(\ref{d3}). However, as explained in Section \ref{BF-dual-sec}, 
it turns out that matching of the sphere free-energy is sensitive to the finite shifts in the CS level appearing in the duality map. Essentially, this is because the pure Chern-Simons theory has a non-trivial partition function on $S^3$, which is known exactly \cite{Witten:1988hf}. In the large $N$ limit, the free-energy of the CS vector models has a term of order $N^2$ coming from the pure 
CS contribution, and a term of order $N$ coming from planar diagrams with a fundamental matter loop. To order $N^2$, the duality is satisfied thanks to the level/rank duality of CS theory. To the order $N$, one obtains a non-trivial duality relation (\ref{bf-duality}) which includes a term that depends on the planar CS free energy, arising 
from the order-one shifts of the level in (\ref{d1})-(\ref{d3}). 

In Section \ref{susy-dual}, we first discuss a closely related duality \cite{Benini:2011mf} for a ${\cal N}=2$ theory with a single fundamental chiral field 
coupled to CS gauge theory. At large $N$, the duality relation to order $N$ involves an analogous mixing with the pure CS term as in the non-SUSY case, and we use localization to 
derive the large $N$ free energy and verify the duality explicitly. We then move on in Section \ref{BF-dual-sec} to the discussion of the non-supersymmetric duality, 
and use it to derive constraints on the $\lambda\rightarrow 1$ strong coupling limit of the scalar and fermion free energies. In Section \ref{relSusy}, we argue that 
in the large $N$ limit the supersymmetric and non-supersymmetric free energies are related in a simple way, so that the scalar and fermion free energies are determined in terms of a single unknown function of `t Hooft coupling. The boson/fermion duality implies that this function should satisfy a certain functional equation, and in (\ref{Bsc-sol}), (\ref{Bfer-sol}) we obtain an explicit solution of this equation satisfying all the relevant weak coupling and strong coupling ``boundary conditions". This is one of the main results of this note. We do not prove that this solution is unique, but it appears to be non-trivial that a solution of the duality relation with the correct behavior does exist. It would be very interesting to perform an explicit perturbative calculation of the free energy on $S^3$ and compare to the perturbative expansions (\ref{bsbf-pert}).   

In Section \ref{wline-sec}, we then study a representation of the sphere free energy in terms of a quantum-mechanical path integral for a particle moving on $S^3$. 
It is well-known that one-loop effective actions in quantum field theory may be expressed 
in terms of worldline path-integrals, 
see e.g. \cite{Strassler:1992zr, Schubert:2001he, Bastianelli:2005rc, Bastianelli:2006rx}. While in this note we focus on $S^3$, the worldline approach works in principle on any manifold, and it would be interesting to generalize our calculations to other backgrounds. For instance, a calculation of the free energy on the thermal hyperbolic 
space $S^1\times H^2$ \cite{Casini:2010kt, Klebanov:2011uf} with temperature $T=1/(2\pi)$ would give an alternative way to obtain $F$, and for $T=1/(2\pi q)$ it would yield predictions for the Renyi entropies $S_q$ of these models.
  
In Section \ref{free-heat}, we first review the calculation of the free-field sphere free energies using the worldline approach (which is essentially equivalent 
to the heat-kernel method), in particular rederiving the known values \cite{Klebanov:2011gs} of $F$ 
for free conformal scalar and fermions. We then move on to the interacting CS vector models in Section \ref{inter-wline}. Through the worldline representation, 
the free energy in the large $N$ limit can be related to the computation of Wilson loop expectation values in pure CS theory on $S^3$, and a path-integral over closed 
loops using the worldline action. The topological nature of CS theory suggests that this may be computed exactly in terms of multiply wound circular Wilson loops (which are 
the saddle points of the worldline action). While this seems plausible, we do not prove it rigorously here. Nevertheless, under this assumption we show that, in the 
supersymmetric theory, the worldline approach precisely reproduces the free-energy computed via localization in Section \ref{susy-dual}. In the non-supersymmetric theories, 
we observe that a result consistent with the duality, and precisely coinciding with the solution constructed in Section \ref{solmap}, can be obtained using a peculiar
framing prescription for Wilson loop expectation values, which remains to be understood. In Section \ref{massive-sec}, we conclude by applying the worldline method in the 
presence of mass parameters. The small mass expansion encodes integrated correlation functions of the $\bar\phi\phi$ and $\bar\psi \psi$ operators, and we find that 
2-point and 3-point functions, which are fixed up to an overall constant by conformal symmetry, have a structure in agreement (modulo fixing certain contact terms in the case of 3-point functions) with known results \cite{Aharony:2012nh,GurAri:2012is} obtained using light-cone gauge in flat space.

\section{${\cal N}=2$ supersymmetric duality}
\label{susy-dual}

Let us start by discussing a close supersymmetric analog to the boson/fermion duality. Consider the theory of a single ${\cal N}=2$ chiral superfield in the fundamental of $U(N)$ coupled to the Chern-Simons gauge field at level $k$. After eliminating all auxiliary fields, the action reads (we follow the conventions 
in \cite{Jain:2012qi})
\begin{equation}
\begin{aligned}
&S=S_{CS_{N,k}}+\int d^3 x\left[D_{\mu}\bar\phi D^{\mu}\phi+\bar\psi \gamma^{\mu}D_{\mu}\psi
+\frac{4\pi}{k}(\bar\phi \phi)(\bar\psi \psi)+\frac{2\pi}{k}(\bar\psi \phi)(\bar\phi \psi)+
\frac{4\pi^2}{k^2} (\bar\phi\phi)^3\right]
\label{Ssusy}
\end{aligned}
\end{equation}
where the pure Chern-Simons action is
\begin{equation}
S_{CS_{N,k}} =\frac{ik}{4\pi}\int d^3x \epsilon^{\mu\nu\rho} \tr (A_{\mu} \partial_\nu A_\rho - \frac{2i}{3} A_\mu A_\nu A_\rho)\,.
\end{equation} 
where $A_{\mu}=A_{\mu}^a T^a$, with $T^a$ generators of $U(N)$ in the fundamental, 
and the trace is normalized in the canonical way, $\tr (T^aT^b)=\frac{1}{2}\delta^{ab}$. 

This theory satisfies a ``chiral" version  \cite{Benini:2011mf} of the Giveon-Kutasov duality \cite{Giveon:2008zn}. 
It is expected to be self-dual under the duality map (see also \cite{Gur-Ari:2015pca}) 
\begin{equation}
U(N)_k ~~~\leftrightarrow ~~~ U(k+\frac{1}{2}-N)_{-k}
\label{duality}
\end{equation}
Note that here $k$ is taken to be half-integer because the models involve a single fundamental fermion. In particular, this ensures that the right-hand side of 
(\ref{duality}) makes sense (here and below, we will assume for simplicity of notations that $k$ is positive). Non-trivial tests of this duality map 
can be obtained by computing the sphere partition function on both sides. Using supersymmetric localization, this can be in principle derived for any $N$ and $k$. Here 
we will be interested in the structure of the duality at large $N$. In the $U(N)_k$ theory, the free energy in the large $N$ `t Hooft limit has the 
expansion
\begin{equation}
F_{{\cal N}=2} = N^2 F_{\rm CS}(\lambda)+ N B_{\rm susy}(\lambda)+\ldots
\label{susy-lN}
\end{equation}
where $\lambda = N/k$ is the 't Hooft coupling. In this expansion, $F_{\rm CS}(\lambda)$ is the planar free energy in pure CS theory, and $B_{\rm susy}(\lambda)$ 
is the contribution to the free energy coming from planar vacuum diagrams with matter loops. The explicit form of $F_{\rm CS}(\lambda)$ is derived in the appendix 
starting from the exact result at finite $N$ and $k$ \cite{Witten:1988hf}, and reads 
\begin{equation}
F_{\rm CS}(\lambda)=\frac{1}{8\pi^2\lambda^2}\left(2\zeta(3)-\mbox{Li}_3(e^{2i\pi\lambda})-\mbox{Li}_3(e^{-2i\pi\lambda})\right)\,.
\label{ACS}
\end{equation}
Comparing with the free energy of the dual theory with $U(k+\frac{1}{2}-N)_{-k}$ gauge group, one finds that at large $N$ the duality is satisfied provided 
\begin{eqnarray}
&&\lambda^2 F_{\rm CS}(\lambda)-(1-\lambda)^2 F_{\rm CS}(1-\lambda) =0\label{CS-duality}\\
&&\lambda B_{\rm susy}(\lambda)-(1-\lambda)B_{\rm susy}(1-\lambda) = (1-\lambda)F_{\rm CS}(1-\lambda)+\frac{1}{2}(1-\lambda)^2 F'_{\rm CS}(1-\lambda)
\label{AB-duality}
\end{eqnarray}
It is easy to see that the identity in the first line is satisfied by (\ref{ACS}), and it just follows from level-rank duality of pure CS theory. The identity in 
the second line is less trivial. It does not have a form of a simple self-duality of $B_{\rm susy}(\lambda)$ under $\lambda \leftrightarrow 1-\lambda$: there is 
a non-trivial right-hand side that essentially comes from the order 1 shift in the rank on the right-hand side of (\ref{AB-duality}), and the fact that pure CS theory has a non-trivial sphere free energy. A similar effect is at play in the non-supersymmetric boson-fermion duality discussed below. Using (\ref{ACS}), we can write (\ref{AB-duality}) more explicitly as
\begin{equation}
\lambda B_{\rm susy}(\lambda)-(1-\lambda)B_{\rm susy}(1-\lambda)=
{i\over 8\pi} \left (\mbox{Li}_2(e^{2i\pi\lambda})-\mbox{Li}_2(e^{-2i\pi\lambda})\right)\,.
\label{B-duality}
\end{equation}

We can check explicity (\ref{B-duality}) by computing $B_{\rm susy}(\lambda)$ using localization. The 
exact result for the partition function is given by (see e.g. \cite{Marino:2011nm, Willett:2016adv} for reviews)
\begin{equation}
\begin{aligned}
Z_{S^3} ={1 \over N!} \int \left( \prod_{i = 1}^{N} { d u_i \over 2 \pi} e^{ i k { u_i^2 \over 4 \pi} } \right) \left( \prod_{i < j}^{N} 4 \sinh^2 \left[ {u_i - u_j  \over 2} \right] \right) \prod_{i = 1}^{N} e^{ \ell(1 - \Delta + i {u_i \over 2 \pi})}
\end{aligned}
\end{equation}
where
\begin{equation}
\ell(z) = - z \log\left(1 - e^{2 \pi i z} \right) + { i \over 2} \left( \pi z^2 + {1 \over \pi} \text{Li}_2 ( e^{2 \pi i z} ) \right) - {i \pi \over 12}\,.
\end{equation}

In the large $N$ limit, the distribution of the eigenvalues $u_i$ is determined by the pure CS theory (this is because matter is in the fundamental), 
and one can use the eigenvalue density for the Chern-Simons matrix model \cite{Marino:2011nm} given by
\begin{equation}
\begin{aligned}
&\rho_0 (u)= {1\over \pi t} \tan^{-1} \left [ {\sqrt{e^t- \cosh^2 (u/2)}\over \cosh (u/2)}\right ]\,,\qquad -a(t)<u<a(t)\\
&a(t)= 2 \cosh^{-1} \exp(t/2)
\end{aligned}
\label{density}
\end{equation}
and the coupling $t$ is related to the CS 't Hooft coupling by analytic continuation, $t=2\pi i \lambda$. Then, one can see that the large $N$ 
free energy $F_{{\cal N}=2}=-\log Z$ has the expansion (\ref{susy-lN}), with 
\begin{equation}
B_{\rm susy}(t) = -\int_{-a(t)}^{a(t)} du \rho_0 (u)\ell(1-\Delta+i \frac{u}{2\pi})\,.
\end{equation}
To leading order in the large $N$ limit, the R-charge $\Delta$ is equal to the free field value $\Delta=\frac{1}{2}+O(1/N)$, and one gets
\begin{equation}
B_{\rm susy}(t) = -\int_{-a(t)}^{a(t)} du \rho_0 (u)\ell(\frac{1}{2}+i \frac{u}{2\pi})=\frac{1}{2}\int_{-a(t)}^{a(t)} du \rho_0 (u)\log(2\cosh(\frac{u}{2}))
\label{B-int}
\end{equation}
where we have used the fact that the density is symmetric in $u\rightarrow -u$ and the identity $\ell(1/2+i x)+\ell(1/2-i x) = -\log(2\cosh(\pi x))$. This 
integral can be evaluated as follows. First, it is convenient to define the function
\begin{equation}
b(t)=t B_{\rm susy}(t)\,.
\end{equation}
Taking the derivative with respect to $t$, one gets\footnote{The derivative acting on the endpoints of the integral does not contribute since the eigenvalue density vanishes at the endpoints.}
\begin{equation}
\begin{aligned}
&\frac{db}{dt}=\int_{-a}^a du\frac{\cosh{u\over 2}}{4\pi\sqrt{e^t-\cosh^2{u\over 2}}}\, \log (2 \cosh (u/2))
=\int_{-\sqrt{e^t-1}}^{\sqrt{e^t-1}}dy\frac{\log(2\sqrt{1+y^2})}{2\pi\sqrt{e^t-1-y^2}}
=\frac{1}{2}\log(1+e^{\frac{t}{2}})\,,
\label{bEval}
\end{aligned}
\end{equation}
where in the second step we have used the change of variables $x=2 \sinh^{-1} y$. Integrating in $t$ with boundary condition $b(0)=0$, we get the result
\begin{equation}
B_{\rm susy}(t)=\frac{1}{2t}\int_0^t dt' \log(1+e^{\frac{t'}{2}})=-\frac{\pi^2}{12t}-\frac{1}{t}{\rm Li}_2(-e^{\frac{t}{2}})\,.
\end{equation}
Analytically continuing to real CS 't Hooft coupling by sending $t  \rightarrow 2\pi i \lambda$, we obtain (disregarding a purely imaginary part)
\begin{equation}
B_{\rm susy}(\lambda)=\frac{i}{4\pi\lambda}\left({\rm Li}_2(-e^{i\pi\lambda})-{\rm Li}_2(-e^{-i\pi\lambda})\right)\,.
\label{Bsu}
\end{equation}
At weak coupling, this function has the expansion
\begin{equation}
B_{\rm susy}=\frac{\log 2}{2}-\frac{\pi ^2 \lambda ^2}{48}-\frac{\pi ^4 \lambda ^4}{1920}-\frac{\pi ^6 \lambda ^6}{40320}+\ldots 
\label{Bsusy-lam0}
\end{equation}
and at ``strong" coupling
\begin{equation}
B_{\rm susy} \simeq -\frac{1}{2}(1-\lambda) \log(1-\lambda) \,,\qquad \lambda \rightarrow 1\,.
\end{equation}
The leading term at weak coupling is the well-known contribution of a free chiral superfield, and it is the sum of the free 
scalar and free fermion contributions (see \cite{Klebanov:2011gs} and eq. (\ref{conf-sc}), (\ref{conf-fer}) below). 

Having the exact result (\ref{Bsu}), one can now explicitly verify that (\ref{B-duality}) is indeed satisfied. This is readily seen using the polylogarithm identity
\begin{equation}
{\rm Li}_{\nu}(z^2)=2^{\nu-1}\left({\rm Li}_{\nu}(z)+{\rm Li}_{\nu}(-z)\right)\,.
\end{equation}

Finally, let us note that it is instructive to rewrite (\ref{B-int}) in the following way. Using 
\begin{equation}
\log(2\cosh(\frac{u}{2}))=-\frac{u}{2}-\sum_{n=1}^{\infty}\frac{(-1)^n}{n}e^{n u}
\end{equation}
we see that\footnote{The linear term in $u$ drops out because the density is symmetric in $u\rightarrow -u$.}
\begin{equation}
B_{\rm susy}(t)  = -\sum_{n=1}^{\infty}\frac{(-1)^n}{2n}\int_{-a(t)}^{a(t)} du \rho_0 (u)  e^{n u}\,.
\end{equation}
But this eigenvalue integral just represents the expectation value of a circular Wilson loop, multiply wrapped $n$ times, in pure CS theory. 
In other words, we can write
\begin{equation}
B_{\rm susy} = -\sum_{n=1}^{\infty} \frac{(-1)^n}{2n} \langle W_n \rangle_{\rm CS}
\label{Bsusy-W}
\end{equation}  
where $W_n$ denotes the multiply wrapped circular Wilson loop, and the average is in pure CS theory in the planar limit.\footnote{We define $W_n$ so that 
the perturbative expansion is $\langle W_n\rangle = 1+O(t)$, i.e. we include a factor of $1/N$ in front of the trace, 
$W=\frac{1}{N}\tr {\cal P} e^{i\oint A}$.} 
Below we show that this form of the result can be recovered from a worldline representation of the free energy.  

\section{Non-supersymmetric boson/fermion duality}
\label{BF-dual-sec}

The non-supersymmetric boson/fermion duality relates the ``critical scalar" theory coupled to Chern-Simons
\begin{equation}
    S_{\rm crit.sc}=S_{CS_{N,k}}+\int d^3x \left( D_{\mu}\bar{\phi} D^{\mu}\phi +\frac{\lambda_4}{4N}  (\bar\phi \phi)^2\right)\,,
\label{CS-boson}
\end{equation}
where it is assumed that we flow to the IR fixed point reached by adding the relevant quartic interaction 
(and tuning the mass to zero), and the fermion theory 
\begin{equation}
    S_{\rm fer}=S_{CS_{N,k}}+\int d^3x \bar{\psi} \slashed{D} \psi\,.
\label{CS-fermion}
\end{equation}
The rank and level of the two theories are related according to the duality map \cite{Aharony:2012nh,Aharony:2015mjs}
\begin{equation}
U(N)_{k} ~~ {\rm Crit.~Scalar}\quad \leftrightarrow \quad 
U(k)_{-N+1/2} ~~{\rm Fermion}\,,
\label{lev-rank-2}
\end{equation} 
Note that in the large $N$ limit, to the order we work, we do not distinguish between versions of the duality (\ref{d1})-(\ref{d3}) involving $SU$ or $U$ groups, and 
also we are not sensitive to the whether the $U(1)$ level is $k$ or $k+N$, and therefore we will write the duality as (\ref{lev-rank-2}). Here we have 
assumed the convention where $k$ is the CS-level that arises when the theory is regularized with a Yang-Mills term in the UV, and the 't~Hooft coupling is 
given by $\lambda=N/(N+k)$. We can also write the duality in terms of the level $\hat k$ in the dimensional reduction \cite{Chen:1992ee} conventions 
\begin{equation}
U(N)_{\hat k} ~~ {\rm Crit.~Scalar}\quad \leftrightarrow \quad 
U(\hat{k}-N)_{-\hat{k}+1/2} ~~{\rm Fermion}\,.
\label{lev-rank-1}
\end{equation}
In this case there is no shift of the level at one-loop (or equivalently $\hat{k}$ may be 
viewed as a renormalized level, $\hat{k} = k+N$, that takes into account 
the one-loop shift by $N$), and the 't Hooft coupling 
is $\lambda = N/\hat{k}$ ($\hat{k}>N$). 

We want to study the implications of (\ref{lev-rank-2}) at the level of the three-sphere free energy at large $N$. First, we note that the value 
of the sphere free energy at the IR fixed point of (\ref{CS-boson}) is not affected by the quartic interaction to order $N$, because that interaction 
can be viewed as the analog of a ``double-trace" deformation, and it only affects the free energy to order $N^0$,\footnote{In an AdS dual description, it corresponds 
to changing the boundary condition to the bulk field dual to $\bar\phi \phi$, and this does not change the bulk free energy at tree level.} see also 
Section \ref{critTh} below. In other words, 
the critical scalar free energy, up to to order $N$, is the same as the one of the ``regular" scalar theory
\begin{equation}
    S_{\rm sc}=S_{CS_{N,k}}+\int d^3x  D_{\mu}\bar{\phi} D^{\mu}\phi \,.
\label{CSsc}
\end{equation}
This is the theory that appears in the alternative 
version of the duality, where one maps the ``regular CS-scalar" to the critical (Gross-Neveu) 
CS-fermion theory. To be more precise, in such regular scalar theory one 
should also add the classically marginal $\frac{\lambda_6}{N^2}  (\bar\phi \phi)^3$ coupling 
(and, similarly, one also has a classically marginal coupling $\sim (\bar\psi\psi)^3$ in 
the critical fermion theory, since $\bar\psi\psi$ has dimension $1+O(1/N)$ at the UV fixed point). 
However, such coupling is exactly marginal at 
large $N$, so it again does not affect the value of the sphere free energy in the planar limit. 
Equivalently, we can see this by noting that this would be a ``triple-trace" term, which also implies that it does 
not affect the sphere free energy to the order $N$. 

To summarize, at the level 
of the CFT sphere free energy in the planar limit, we can then consider the theories (\ref{CSsc}) and (\ref{CS-fermion}) 
and ignore the difference with the 
critical versions. Note that this is no longer true in the presence of mass parameters 
(the massive case will be discussed in more detail in Section \ref{massive-sec} below). It 
is also not true in the case of the thermal free energy on $S^1\times \mathbb{R}^2$ (or $S^1\times S^2$), 
which even in the planar limit is different for critical and regular theories (and it does depend on $\lambda_6$) \cite{Jain:2012qi,Aharony:2012ns}. 

To compute the sphere partition functions, one places the theories (\ref{CSsc}) and (\ref{CS-fermion}) on the round $S^3$ and computes the path-integrals
\begin{equation}
\begin{aligned}
&Z_{\rm sc} = \int {\cal D}A {\cal D}\bar\phi {\cal D}\phi\, e^{-S_{CS}-\int d^3x\sqrt{g} \left(D_{\mu}\bar{\phi} D^{\mu}\phi +\frac{3}{4r^2}\bar\phi\phi\right)}
= \int {\cal D}A\, e^{-S_{CS}-\log\det \left(-D_{\mu}D^{\mu}+\frac{3}{4r^2}\right)}\,,\\
&Z_{\rm fer}=\int {\cal D}A {\cal D}\bar\psi {\cal D}\psi\,  e^{-S_{CS}-\int d^3x\sqrt{g} \bar{\psi} \slashed{D} \psi} 
= \int  {\cal D}A\, e^{-S_{CS} + \log\det \left(\slashed{D}\right)}\,,
\label{Zbf}
\end{aligned}
\end{equation}
where the term $3/(4r^2)\bar\phi\phi$ in the scalar theory is the conformal coupling to the sphere curvature. 
We will set the radius $r=1$ in what follows.

The free energies of scalar and fermion theories have the following large $N$ expansions
\begin{equation}
F_{\rm sc} = N^2 F_{\rm CS}(\lambda)+N B_{\rm sc}(\lambda)+O(N^0)
\label{Fsc-lN}
\end{equation}
and 
\begin{equation}
F_{\rm fer} = N^2 F_{\rm CS}(\lambda)+N B_{\rm fer}(\lambda)+O(N^0)\,,
\label{Ffer-lN}
\end{equation}
where $N,\lambda$ are the rank and 't Hooft coupling in the respective theories. Then, one finds that the duality map 
(\ref{lev-rank-2}), or (\ref{lev-rank-1}), expanded at large $N$, holds to order $N$ provided that\footnote{I am grateful 
to I. Klebanov for collaboration on the derivation of this large $N$ duality relation.} 
\begin{equation}
\label{bf-duality}
\lambda B_{\rm sc}(\lambda) -(1-\lambda) B_{\rm fer}(1-\lambda)= \frac{1}{2}  (1-\lambda)^3 F'_{\rm CS}(1-\lambda)\,,
\end{equation}
and the dominant term of order $N^2$ is the same as (\ref{CS-duality}) and works automatically 
as in the SUSY case by virtue of level/rank duality in pure CS theory. 
As in (\ref{AB-duality}), 
we see that there is a non-zero right-hand side in (\ref{bf-duality}), which is due to the half-integer shift 
in the CS level on the fermionic side of (\ref{lev-rank-2}). 

At weak coupling, using the known values of $F$ for free conformal fields \cite{Klebanov:2011gs}, we have
\begin{equation}
\begin{aligned}
&B_{\rm sc}(\lambda) =  \frac{\log 2}{4} - \frac{3\zeta(3)}{8 \pi^2} + O(\lambda^2)\,,\\
&B_{\rm fer}(\lambda)=  \frac{\log 2}{4} + \frac{3\zeta(3)}{8 \pi^2} + O(\lambda^2)\,.
\label{lam0}
\end{aligned}
\end{equation}
Then, the duality relation (\ref{bf-duality}) together with the explicit form of the CS free energy (\ref{ACS}), predicts the following behavior 
as $\lambda\rightarrow 1$ 
\begin{equation}
\begin{aligned}
&B_{\rm sc} (\lambda\rightarrow 1) = (1-\lambda) \left ( \frac{\log 2}{4} + \frac{3\zeta(3)}{8 \pi^2} \right ) + O((1-\lambda)^2) \,, \\
&B_{\rm fer}(\lambda\rightarrow 1) =  - \frac 1 2 (1-\lambda) \log (1-\lambda)+ O(1-\lambda) \,.
\label{lam1}
\end{aligned}
\end{equation}
The logarithmic term in the strong coupling behavior of $B_{\rm fer}$ comes from $F'_{\rm CS}(\lambda)$, 
and thus crucially depends on the mixing with the Chern-Simons term.

\subsection{Relation to supersymmetric free energy}
\label{relSusy}

In this section we derive a relation between the sphere free energy of the ${\cal N}=2$ SUSY theory (\ref{Ssusy}) and 
the non-supersymmetric CS-scalar and CS-fermion theories. We can actually consider a more general theory 
with fundamental bosons and fermions coupled to $U(N)_k$ CS theory
\begin{equation}
\begin{aligned}
&S=S_{CS_{N,k}}+\int d^3 x\left[D_{\mu}\bar\phi D^{\mu}\phi+\bar\psi \gamma^{\mu}D_{\mu}\psi\right.\\ 
&~~~~~~~~~~~~~~\left.+\frac{g_4}{N}(\bar\phi \phi)(\bar\psi \psi)+\frac{g_6}{N^2} (\bar\phi\phi)^3+\frac{g_4^{'}}{N}(\bar\psi \phi)(\bar\phi \psi)
+\frac{g_4''}{N}((\bar\psi \phi)(\bar\psi\phi)+
(\bar\phi \psi)(\bar\phi\psi))\right]
\label{SBF}
\end{aligned}
\end{equation} 
where in the supersymmetric theory $g_4, g_6, g_4', g_4''$ have some fixed values that 
can be read-off from (\ref{Ssusy}), and are of order $\lambda$ in the large $N$ limit. 

One can see that the only couplings that could contribute to the computation of the free energy 
of the model (\ref{SBF}) to order $N$ are $g_4$ and $g_6$ (a similar fact holds for the thermal free energy 
\cite{Jain:2012qi, Aharony:2012ns}). In fact, in the conformal case (when masses are tuned to zero), one can see that 
the sphere free energy to order $N$ is independent of $g_4$ and $g_6$ as well. This is because these couplings are double-trace 
and triple-trace deformations and so will affect the free energy starting 
at order $N^0$ (note that this is not true in the case of the thermal free energy, 
which does depend on $g_4$ and $g_6$ to leading order in $N$ \cite{Jain:2012qi, Aharony:2012ns}). 
Then, at large $N$ the free energy of the ${\cal N}=2$ model may be obtained from 
\begin{eqnarray}
F_{{\cal N}=2} &\approx & -\log\int {\cal D}A {\cal D}\bar\phi {\cal D}\phi {\cal D}\bar\psi {\cal D}\psi \,
e^{-S_{CS}-\int d^3x\sqrt{g} \left(D_{\mu}\bar{\phi} D^{\mu}\phi +\frac{3}{4r^2}\bar\phi\phi+\bar{\psi} \slashed{D} \psi\right)}\cr 
&=& -\log \int {\cal D}A e^{-S_{CS} -\log\det \left(-D_{\mu}D^{\mu}+\frac{3}{4r^2}\right)+ \log\det \left(\slashed{D}\right)}\,.
\end{eqnarray}
To the order $N$, the fermion and boson contributions decouple and we have
\begin{equation}
F_{{\cal N}=2}=N^2 F_{\rm CS}(\lambda)+N (B_{\rm sc}(\lambda)+B_{\rm fer}(\lambda))+O(N^0)\,,
\label{Fsusy}
\end{equation}
or in other words\footnote{I thank I. Klebanov for a discussion on the relation (\ref{Bsu-to-Bsf}) between supersymmetric and non-supersymmetric 
free energies.} 
\begin{equation}
B_{\rm susy}(\lambda) = B_{\rm sc}(\lambda)+B_{\rm fer}(\lambda)
\label{Bsu-to-Bsf}
\end{equation}
with $B_{\rm susy}(\lambda)$ given in (\ref{Bsu}). This implies that the scalar and fermion free energy are determined by a single 
unknown function of `t Hooft coupling
\begin{equation}
\begin{aligned}
B_{\rm sc}(\lambda) = \frac{1}{2}B_{\rm susy}(\lambda)+\frac{g(\lambda)}{\lambda}\,,\qquad 
B_{\rm fer}(\lambda) = \frac{1}{2}B_{\rm susy}(\lambda)-\frac{g(\lambda)}{\lambda} \,. 
\label{bf-susy}
\end{aligned}
\end{equation}

\subsection{A solution to the duality map}
\label{solmap}

Using the relation (\ref{bf-susy}), the supersymmetric duality relation (\ref{AB-duality}), and the boson/fermion duality 
(\ref{bf-duality}), one can obtain the following functional equation that $g(\lambda)$ should satisfy
\begin{equation}
g(\lambda)+g(1-\lambda) = \frac{1}{4}(1-2\lambda)(1-\lambda)^2 F_{\rm CS}'(1-\lambda) -\frac{1}{2}(1-\lambda)F_{\rm CS}(\lambda)\,.
\label{g-eq}
\end{equation}
A consistency check of this equation is that the right-hand side is indeed symmetric 
under $\lambda \leftrightarrow 1-\lambda$. This is not obvious, but follows 
from the explicit form of $F_{\rm CS}(\lambda)$. 

An explicit solution to (\ref{g-eq}) satisfying the ``boundary conditions" (\ref{lam0}) and (\ref{lam1}) is given by 
\begin{equation}
\begin{aligned}
&g(\lambda) = \frac{i \lambda ^2\left(\text{Li}_2\left(e^{-2 i \pi  \lambda }\right)-\text{Li}_2\left(e^{2 i \pi  \lambda }\right)\right)}{{16 \pi }}
+\frac{\lambda  \left(\text{Li}_3\left(e^{-2 i \pi  \lambda }\right)+\text{Li}_3\left(e^{2 i \pi  \lambda }\right)-2 \zeta (3)\right)}{8 \pi ^2}\\
&~~~~~~
-\frac{3 i \left(\text{Li}_4\left(e^{-2 i \pi  \lambda }\right)-\text{Li}_4\left(e^{2 i \pi  \lambda }\right)\right)}{32 \pi ^3}
\end{aligned}
\label{gsol}
\end{equation}
Plugging this solution into (\ref{bf-susy}) one then obtains the scalar and fermion free-energies 
\begin{equation}
\begin{aligned}
&B_{\rm sc}(\lambda) =\frac{i}{8\pi\lambda}\left({\rm Li}_2(-e^{i\pi\lambda})-{\rm Li}_2(-e^{-i\pi\lambda})\right)
-\frac{3 i}{32 \pi ^3\lambda} \left(\text{Li}_4(e^{-2 i \pi  \lambda })-\text{Li}_4(e^{2 i \pi  \lambda })\right)\\
&~~~~~~~~~+ \frac{i \lambda}{16 \pi}\left(\text{Li}_2(e^{-2 i \pi  \lambda })-\text{Li}_2(e^{2 i \pi  \lambda })\right)
+\frac{1}{8 \pi ^2}\left(\text{Li}_3(e^{-2 i \pi  \lambda })+\text{Li}_3(e^{2 i \pi  \lambda })-2 \zeta (3)\right)
\label{Bsc-sol}
\end{aligned}
\end{equation}
\begin{equation}
\begin{aligned}
&B_{\rm fer}(\lambda) =\frac{i}{8\pi\lambda}\left({\rm Li}_2(-e^{i\pi\lambda})-{\rm Li}_2(-e^{-i\pi\lambda})\right)
+\frac{3 i}{32 \pi ^3\lambda} \left(\text{Li}_4(e^{-2 i \pi  \lambda })-\text{Li}_4(e^{2 i \pi  \lambda })\right)\\
&~~~~~~~~~- \frac{i \lambda}{16 \pi}\left(\text{Li}_2(e^{-2 i \pi  \lambda })-\text{Li}_2(e^{2 i \pi  \lambda })\right)
-\frac{1}{8 \pi ^2}\left(\text{Li}_3(e^{-2 i \pi  \lambda })+\text{Li}_3(e^{2 i \pi  \lambda })-2 \zeta (3)\right)
\label{Bfer-sol}
\end{aligned}
\end{equation}
These satisfy the expected boson/fermion duality relation (\ref{bf-duality}), and have the correct behavior at $\lambda \ll 1$ and $\lambda\rightarrow 1$. The 
explicit perturbative expansions are found to be 
\begin{equation}
\begin{aligned}
&B_{\rm sc}(\lambda) = \frac{\log 2}{4}-\frac{3 \zeta (3)}{8 \pi ^2}
-\frac{1}{96} \left(\pi^2+4\right) \lambda ^2
-\frac{\left(\pi^4-16\pi^2\right) \lambda ^4}{3840}
-\frac{\left(3 \pi ^6-32 \pi^4\right) \lambda ^6}{241920}+\ldots \\
&B_{\rm fer}(\lambda) =\frac{\log 2}{4}+\frac{3 \zeta (3)}{8 \pi ^2}
-\frac{1}{96} \left(\pi ^2-4\right) \lambda ^2
-\frac{\left(\pi^4+16\pi^2\right)\lambda ^4}{3840}
-\frac{\left(3\pi ^6+32\pi ^4\right) \lambda ^6}{241920}+\ldots 
\label{bsbf-pert}
\end{aligned}
\end{equation}
and the $\lambda\rightarrow 1$ behavior 
is in agreement with (\ref{lam1}). Note that in the small $\lambda$ expansion (\ref{bsbf-pert}) 
the terms of highest transcendentality (beyond the $\lambda^0$ term) are the same 
in scalar and fermion theories, and are fixed by the supersymmetric result, see (\ref{Bsusy-lam0}). 

While we cannot prove that the solution (\ref{gsol}) is unique, it seems non-trivial that such a relatively simple solution 
with the correct behavior exists. Of course, it would be interesting to check the above perturbative expansions (\ref{bsbf-pert}) 
by a direct Feynman diagram calculation on $S^3$.

\section{Worldline representation of the free energy}
\label{wline-sec}

\subsection{Free scalar and fermion heat kernel as sum over geodesics}
\label{free-heat}

Consider a free massive complex fundamental scalar on $S^3$. The free energy is given by the one-loop determinant
\begin{equation}
F_{\rm free-sc}(m) = \log \det \left(-\nabla^2+m^2+\frac{3}{4}\right)
\end{equation}
where the factor of $3/4$ comes from the conformal coupling to the scalar curvature (we set 
the radius $r=1$ in what follows), so that $m^2=0$ corresponds to a conformal scalar field.

One has 
the following representation of the determinant on a general manifold \cite{Strassler:1992zr, Schubert:2001he, Bastianelli:2005rc, Bastianelli:2006rx}
\begin{equation}
\log\det\left(-\nabla^2+m^2+\xi R\right)=-\int_0^{\infty}\frac{d\beta}{\beta}
\int_
{x^{\mu}(0)=x^{\mu}(1)}\!\!\!\!\!\!\!\!\!\!\!\!\!\!\!\!
{\cal D}x\, \,e^{-\int_0^1 d\tau \left(\frac{1}{4\beta}g_{\mu\nu}(x)\dot x^{\mu}\dot x^{\nu}+\beta (m^2+\xi R)\right)}
\label{WL-path}
\end{equation}
where the path-integral is over closed loops on the manifold, and the worldline action is that 
of a relativistic scalar particle on a curved space with metric with metric $g_{\mu\nu}(x)$. In 
this formula we have allowed for a general coupling to the scalar curvature parameterized by the constant $\xi$. 
A conformally coupled field in 3d corresponds to $\xi = 1/8$ and $m=0$.  

The path-integral over $x(\tau)$ with periodic boundary condition computes the trace of the scalar heat kernel 
at coincident points
\begin{equation}
\int_
{x^{\mu}(0)=x^{\mu}(1)}\!\!\!\!\!\!\!\!\!\!\!\!\!\!\!\!
{\cal D}x\, {\rm tr}{\cal P}\,e^{-\int_0^1 d\tau \left(\frac{1}{4\beta}g_{\mu\nu}(x)\dot x^{\mu}\dot x^{\nu}+\beta (m^2+\xi R)\right)}=\int d^3x\sqrt{g(x)} K(x,x;\beta)\,,
\end{equation}
where $K(x,x';t)$ satisfies the heat equation
\begin{equation}
\left(-\nabla^2_x+m^2+\xi R +\partial_{\beta}\right)K(x,x';\beta)=0\,.
\label{heat}
\end{equation}

In the case of $S^3$, the scalar heat-kernel is known exactly, see e.g. \cite{Camporesi:1990wm}. 
Choosing $\xi =1/8$, so that $\xi R = 3/4$ and $m^2=0$ gives the conformal scalar, 
the heat kernel is given explicitly by 
\begin{equation}
K_{S^3}(x,x';\beta)=K_{S^3}(\sigma(x,x');\beta)=\frac{1}{(4\pi\beta)^{3/2}}\sum_{n=-\infty}^{\infty} \frac{\sigma+2\pi n}{\sin\sigma}\,e^{-\frac{(\sigma+2\pi n)^2}{4\beta}-\beta(m^2-\frac{1}{4})}\,.
\label{Ksc}
\end{equation}
Here $\sigma(x,x')$ is the geodesic distance between $x$ and $x'$, and the sum over $n$ corresponds 
to a sum over geodesics from $x$ to $x'$ which wind $n$ times along a great circle; $n=0$ 
corresponds to the ``direct path" between $x$ and $x'$. 

To obtain the one-loop determinant, we have to take the trace of the scalar heat kernel at 
coincident points. Taking the limit $\sigma\rightarrow 0$ of (\ref{Ksc}) (and using 
a symmetric prescription for the summation over $n$), one finds
\begin{equation}
\int d^3x\sqrt{g(x)} K_{S^3}(\sigma\rightarrow 0 ;\beta) = {\rm vol(S^3)} \frac{e^{-(m^2+\frac{3}{4})\beta}}{(4\pi\beta)^{3/2}}\left(e^{\beta}
+2\sum_{n=1}^{\infty} e^{-\frac{n^2\pi^2}{\beta}}e^{\beta}(1-\frac{2n^2\pi^2}{\beta})\right)
\label{K-coinc}
\end{equation}

This result has a natural interpretation from the point of view of the quantum mechanical 
path-integral (\ref{WL-path}). 
The sum over $n$ corresponds to the sum over the classical saddle points, i.e. the solutions to the 
equation of motion coming from the worldline action
\begin{equation}
\ddot x^{\mu}+\Gamma^{\mu}_{\nu \rho}\dot x^{\nu}\dot x^{\rho}=0\,,
\label{eom}
\end{equation} 
with boundary condition $x^{\mu}(0)=x^{\mu}(1)$, namely closed geodesics. On $S^3$ with the round metric, these are circular loops 
winding $n$ times around an equator of $S^3$.  For each such saddle point, one then computes the path-integral 
over quantum fluctuations around it. In other words, one writes 
\begin{equation}
x(\tau) = x_n (\tau) + \sqrt{\beta}\xi(\tau)
\end{equation}
where $x_n(\tau)$ is the $n$-wrapped geodesic, and $\xi(\tau)$ are quantum fluctuations with periodic boundary conditions. 
The path integral can be then expressed as\footnote{The factor of $\frac{\rm vol(S^3)}{(4\pi\beta)^{3/2}}$ 
comes from the integration over the translational zero mode, which is the 
``base point" for the geodesic.}
\begin{eqnarray}
\int_{x^{\mu}(0)=x^{\mu}(1)}\!\!\!\!\!\!\!\!\!\!\!\!\!\!\!\!
{\cal D}x\, {\rm tr}{\cal P}\,e^{-\int_0^1 d\tau \left(\frac{1}{4\beta}g_{\mu\nu}(x)\dot x^{\mu}\dot x^{\nu}+\beta (m^2+\frac{3}{4})\right)}
&=&{\rm vol(S^3)} \frac{e^{-\beta(m^2+\frac{3}{4})}}{(4\pi\beta)^{3/2}}\sum_{n}\int_{\rm PBC} D\xi e^{-S[x_n+\sqrt{\beta}\xi]}\cr
&=&{\rm vol(S^3)} \frac{e^{-\beta(m^2+\frac{3}{4})}}{(4\pi\beta)^{3/2}}\sum_{n} e^{-S_{\rm cl}(n)}f_n(\beta)
\end{eqnarray}
In (\ref{K-coinc}), the first term in parenthesis is the contribution of the trivial geodesic ($n=0$), and the factor $e^{\beta}$ comes from the path-integral 
over quantum fluctuations. The second term in (\ref{K-coinc}) is then the contribution 
of the non-trivial geodesics, where the factor $e^{-n^2\pi^2/\beta}$ corresponds 
to the classical worldline action evaluated on the $n$-wound geodesic.\footnote{Using the standard 
polar coordinates $ds^2 = d\theta_1^2+\sin^2\theta_1(d\theta_2^2+\sin^2\theta_2 d\phi^2)$, a multiply wound geodesic 
solving (\ref{eom}) is given by $\theta_1=\theta_2=\frac{\pi}{2}$ and $\phi(\tau) = 2\pi n \tau$. The worldline action 
evaluated on this solution is $S = \int_0^1 d\tau \frac{1}{4\beta}\left(\dot\theta_1^2+
 \sin^2\theta_1(\dot\theta_2^2+\sin^2\theta_2 \dot\phi^2)\right) = n^2\pi^2/\beta$.}  
As far as we know, the result 
(\ref{K-coinc}) was not derived in the literature directly from the quantum mechanical path-integral,\footnote{The 
contribution of the $n=0$ geodesic can be seen to match general results 
for the small $\beta$ expansion of 
the heat kernel on general metric obtained directly from the worldline path-integral, 
see e.g. \cite{Bastianelli:2002fv, Bastianelli:2006rx}, but we are 
not aware of a direct calculation of the path-integral around the non-trivial geodesics.}
nevertheless the known exact solution for the heat kernel has the expected structure.

To obtain the one-loop determinant, we have to integrate (\ref{K-coinc}) over the proper time $\beta$. Then we obtain the $S^3$ free energy for 
a free complex scalar field
\begin{eqnarray}
F_{\rm free-sc}(m) &=& \log\det\left(-\nabla^2+m^2+\frac{3}{4}\right)\cr
&=&-\int_0^{\infty}\frac{d\beta}{\beta} {\rm vol(S^3)} \frac{e^{-(m^2+\frac{3}{4})\beta}}{(4\pi\beta)^{3/2}}\left(e^{\beta}
+2\sum_{n=1}^{\infty} e^{-\frac{n^2\pi^2}{\beta}}e^{\beta}(1-\frac{2n^2\pi^2}{\beta})\right)\\
&=&-2\pi^2\left(\frac{(m^2-\frac{1}{4})^{\frac{3}{2}}}{6\pi}+\sum_{n=1}^{\infty}\frac{2n^2\pi^2(\frac{1}{4}-m^2)-1-2\pi n \sqrt{m^2-\frac{1}{4}}}{4\pi^4n^3}\,e^{-2\pi n \sqrt{m^2-\frac{1}{4}}}\right)\nonumber
\label{logDetfree}
\end{eqnarray}
where we have used ${\rm vol}(S^3)=2\pi^2$. Specializing to the conformal case $m=0$, this reduces to 
\begin{eqnarray}
F_{\rm conf-sc} &=& 
\frac{i\pi}{24}-\sum_{n=1}^{\infty} (-1)^{n}\left(\frac{1}{4n}-\frac{1}{2\pi^2 n^3}-\frac{i}{2\pi n^2}\right)\cr 
&=&-\sum_{n=1}^{\infty} (-1)^{n}\left(\frac{1}{4n}-\frac{1}{2\pi^2 n^3}\right)= \frac{\log 2}{4}-\frac{3\zeta(3)}{8\pi^2}\,.
\label{conf-sc}
\end{eqnarray}
This is the expected result for the $F$ coefficient of a free (complex) conformal scalar \cite{Klebanov:2011gs}. Note 
that in this heat kernel approach, this is obtained from a ``sum over geodesics" which appears to be more convergent 
than the sum over eigenvalues of the differential operator \cite{Klebanov:2011gs} (see also \cite{Marino:2011nm}). 
The two approaches should be related by a Poisson resummation. The ``duality" 
between sum over geodesics and spectral sum can be seen more generally on symmetric spaces, see \cite{Camporesi:1990wm}. 

For non-zero mass, the sum (\ref{logDetfree}) can be evaluated explicitly in terms of polylogarithms
\begin{equation}
\begin{aligned}
&F_{\rm free-sc}(m) = -\frac{\pi}{3}\left(m^2-\frac{1}{4}\right)^{\frac{3}{2}}
-\left(m^2-\frac{1}{4}\right)\log\left(1-e^{-2\pi \sqrt{m^2-\frac{1}{4}}}\right)\\
&~~~~~~~~~~~~~~~~~
+\frac{\sqrt{m^2-\frac{1}{4}}}{\pi}\,
{\rm Li}_2(e^{-2\pi \sqrt{m^2-\frac{1}{4}}})
+\frac{1}{2\pi^2} \,
{\rm Li}_3(e^{-2\pi \sqrt{m^2-\frac{1}{4}}})\,.
\end{aligned}
\end{equation}
This agrees with the known result given e.g. in \cite{Giombi:2014xxa}. The small mass expansion reads 
\begin{equation}
F_{\rm free-sc}(m) =\left(\frac{\log 2}{4}-\frac{3 \zeta (3)}{8 \pi ^2}\right)-\frac{\pi^2}{8}m^4+\frac{\pi^2}{12}m^6+(\frac{\pi^2}{8}-\frac{\pi^4}{48})m^8+\ldots 
\label{free-small-m}
\end{equation}
Note that the coefficient of $m^{2n}$ in this expansion is related to the integrated $n$-point 
function of the $j_0 = \bar\phi\phi$ operator in 
the free massless theory. For two and three point functions, corresponding to the $m^4$ and $m^6$ terms, this 
can be checked using the integrals \cite{Cardy:1988cwa, Klebanov:2011gs}:
\begin{align}
&I_{2}(\Delta)=\int \frac{d^{d}xd^{d}y\sqrt{g_{x}}\sqrt{g_{y}}}{s(x,y)^{2\Delta}}= (2 r)^{2 (d-\Delta )}\frac{2^{1-d} \pi ^{d+\frac{1}{2}} \Gamma \left(\frac{d}{2}-\Delta \right)}{\Gamma \left(\frac{d+1}{2}\right) \Gamma (d-\Delta )}\,, \notag\\
& I_{3}(\Delta) =\int \frac{d^{d}xd^{d}yd^{d}z\sqrt{g_{x}}\sqrt{g_{y}}\sqrt{g_{z}}}{[s(x,y)s(y,z)s(z,x)]^{\Delta}}= r^{3(d-\Delta)} \frac{8\pi^{\frac{3(1+d)}{2}}\Gamma(d-\frac{3\Delta}{2})}{\Gamma(d)\Gamma(\frac{1+d-\Delta}{2})^{3}}\,,
\label{I2I3}
\end{align}
where $s(x,y)$ is the chordal distance (recall that correlation functions of primaries on the sphere can be 
obtained from those in flat space by a conformal transformation, which essentially 
amounts to replacing $|x-y|\rightarrow s(x,y)$.) Note that the term of order $m^2$ is missing in (\ref{free-small-m}) 
as expected from conformal invariance, as one-point functions on the sphere vanish in a conformal theory. 

A similar analysis can be carried out for a free Dirac fermion. The free energy can be obtained from the 
one-loop determinant of the iterated Dirac operator, which admits a worldline representation in terms of 
${\cal N}=1$ supersymmetric quantum mechanics 
\begin{equation}
\begin{aligned}
&F_{\rm free-fer}(m)=-\frac{1}{2}\log\det\left (\slash \!\!\!\!\nabla^2 -m^2\right)\\
&=\frac{1}{2}\int_0^{\infty}\frac{d\beta}{\beta}\int_{PBC}
{\cal D}x\, \int_{ABC} {\cal D}\psi \,e^{-\int_0^1 d\tau \left(\frac{1}{4\beta}g_{\mu\nu}(x)\dot x^{\mu}\dot x^{\nu}+\frac{1}{2}\psi^{\mu}D_{\tau}\psi^{\nu}g_{\mu\nu}(x)+\beta m^2\right)}
\label{Ffree-fer}
\end{aligned}
\end{equation}
where the path integral is with periodic boundary conditions (PBC) on $x^{\mu}$ 
and antiperiodic (ABC) on $\psi^{\mu}$, and $D_{\tau}\psi^{\nu}=\dot \psi^{\nu}+\dot x^{\rho}\Gamma^{\nu}_{\rho\sigma}\psi^{\sigma}$.
This path-integral computes the trace of the spinor heat kernel at coincident points, which satisfies the heat equation 
for the squared Dirac operator. The explicit expression for the spinor heat kernel on $S^3$ can be found in \cite{Camporesi:1992tm}. 
It is given by the $2\times 2$ matrix (i.e. it is a bispinor) 
\begin{equation}
K_{\rm fer}(\sigma(x,x');\beta)=\frac{e^{-\beta m^2}}{(4\pi\beta)^{3/2}}
\left(\cos\frac{\sigma}{2}+i \vec{n}\cdot \vec{\tau} \sin\frac{\sigma}{2} \right)
\frac{1}{\sin\sigma}\sum_{n=-\infty}^{\infty}(-1)^n \left(\sigma+2\pi n-\beta \tan\frac{\sigma}{2}\right)e^{-\frac{(\sigma+2\pi n)^2}{4\beta}}\,,
\end{equation}
where $\vec\tau$ are Pauli matrices and $\vec{n}$ a unit 3-vector. 
Taking the trace over spinor indices and the limit of coincident points, one obtains
\begin{equation}
\begin{aligned}
&\int d^3x\sqrt{g} {\rm tr}K_{\rm fer}(\sigma(x,x);\beta)
&={\rm vol}(S^3) \frac{e^{-\beta m^2}}{(4\pi \beta)^{\frac{3}{2}}}\left(2-\beta
+2\sum_{n=1}^{\infty} (-1)^n e^{-\frac{n^2\pi^2}{\beta}}\left(2-\beta-\frac{4n^2\pi^2}{\beta}\right)\right)
\end{aligned}
\label{Kf-coinc}
\end{equation}
As in the scalar case above, this has an interpretation as a sum over the classical solutions 
of the worldline action, i.e. the multiply wound geodesics on the sphere. 
Integrating (\ref{K-coinc}) over the proper time, one gets the free energy for a free massive Dirac fermion on $S^3$
\begin{equation}
\begin{aligned}
&F_{\rm free-fer}(m) =-\frac{1}{2}\log\det\left (\slash \!\!\!\!\nabla^2 -m^2\right) 
=\frac{1}{2}\int_0^{\infty} \frac{d\beta}{\beta} \int d^3x\sqrt{g} {\rm tr}K_{\rm fer}(\sigma(x,x);\beta)\\
&={\rm vol}(S^3)\left(\frac{m(4m^2+3)}{24\pi}-\sum_{n=1}^{\infty} (-1)^n e^{-2n\pi m }\frac{n^2\pi^2(4m^2+1)+4n\pi m+2}{8\pi^4 n^3}\right)
\label{Ffer-m}
\end{aligned}
\end{equation}
In the conformal case, this gives 
\begin{equation}
F_{\rm conf-fer} = -\sum_{n=1}^{\infty} (-1)^{n}\left(\frac{1}{4n}+\frac{1}{2\pi^2 n^3}\right)
= \frac{\log 2}{4}+\frac{3\zeta(3)}{8\pi^2}\,,
\label{conf-fer}
\end{equation}
in agreement with the known result \cite{Klebanov:2011gs}. For non-zero $m$, the sum (\ref{Ffer-m}) evaluates to  
\begin{equation}
F_{\rm free-fer}(m) = \frac{\pi  m}{12}  \left(4 m^2+3\right)+\frac{1}{4} \left(4 m^2+1\right) \log \left(1+e^{-2 \pi  m}\right)
-\frac{m}{\pi} \text{Li}_2\left(-e^{-2 m \pi }\right)-\frac{1}{2 \pi ^2}\text{Li}_3\left(-e^{-2 m \pi }\right)\,.
\end{equation}
Its small mass expansion reads explicitly
\begin{equation}
F_{\rm free-fer}(m) = \left(\frac{\log 2}{4}+\frac{3\zeta(3)}{8\pi^2}\right)+\frac{\pi^2}{8}m^2 +(\frac{\pi^2}{4}-\frac{\pi^4}{48}) m^4+\ldots 
\end{equation}
The coefficients of this expansion reproduce the integrated $n$-point functions of the $\bar\psi \psi$ operator 
in the free massless Dirac fermion theory. Note that the order $m^3$ does not appear, 
since the 3-point function vanishes by parity symmetry.

\subsection{Chern-Simons interactions and sum over Wilson loops}
\label{inter-wline}

The worldline representation described in the previous section can be generalized to include the coupling 
to the $U(N)$ Chern-Simons gauge theory. 
The sphere free energy of the scalar and fermion theories takes the form (\ref{Zbf}), and hence involves 
computing one-loop determinants in the presence of the gauge field. Let us start with the scalar theory, 
and focus on the massless case for the remaining of this section. The relevant one-loop determinant 
in (\ref{Zbf}) has the worldline representation 
\begin{equation}
\log\det\left(-\nabla^2_A+\frac{3}{4}\right) 
=-\int_0^{\infty}\frac{d\beta}{\beta}\int_
{x^{\mu}(0)=x^{\mu}(1)}\!\!\!\!\!\!\!\!\!\!\!\!\!\!\!\!
{\cal D}x\, {\rm tr}{\cal P}\,
e^{-\int_0^1d\tau \left(\frac{1}{4\beta}g_{\mu\nu}(x)\dot x^{\mu}\dot x^{\nu}+\frac{3}{4}\beta
-i\dot x^{\mu}A_{\mu}(x)\right)}
\label{logDetA}
\end{equation} 
Here the trace is over the $U(N)$ indices and ${\cal P}$ denotes the usual path-ordering. The 
trace ${\rm tr}$ is in the fundamental representation.  
Note that the dependence on the gauge field is through the Wilson loop operator
\begin{equation}
W_{x(\tau)}= \frac{1}{N}{\rm tr}{\cal P} \,e^{i\oint \dot x^{\mu}A_{\mu}}\,.
\end{equation}
Inserting the worldline representation (\ref{logDetA}) into the partition function 
(\ref{Zbf}), we may proceed by first performing the path-integral over the gauge field, 
before the average over closed paths. 
We have
\begin{eqnarray}
Z_{\rm sc} &=&\int {\cal D}A e^{-S_{CS_{N,k}}+N\int_0^{\infty}\frac{d\beta}{\beta}\int
{\cal D}x\, e^{-\int_0^1 d\tau \left(\frac{1}{4\beta}g_{\mu\nu}(x)\dot x^{\mu}\dot x^{\nu}+\frac{3}{4}\beta\right)}W_{x(\tau)}}\\
&=&  Z_{CS_{N,k}} 
\langle e^{N\int_0^{\infty}\frac{d\beta}{\beta}\int
{\cal D}x\, e^{-\int_0^1 d\tau \left(\frac{1}{4\beta}g_{\mu\nu}(x)\dot x^{\mu}\dot x^{\nu}+\frac{3}{4}\beta\right)}W_{x(\tau)}} \rangle_{\rm CS}
\end{eqnarray}
where $\langle \cdots \rangle_{\rm CS}$ denotes the average in pure CS theory, normalized by the CS partition function. 
Using large $N$ 
factorization\footnote{See \cite{Correale:1994mx} for a discussion of the factorization 
of Wilson loop correlators in Chern-Simons theory at large $N$.} 
\begin{equation}
\langle W_{1} W_{2}\cdots W_{k}\rangle \approx  \langle W_{1}\rangle\cdots \langle W_{k}\rangle
\end{equation}
we can then approximate 
\begin{equation}
Z_{\rm sc} \approx Z_{CS_{N,k}}\, e^{N\int_0^{\infty}\frac{d\beta}{\beta}\int
{\cal D}x\, e^{-\int_0^1 d\tau \left(\frac{1}{4\beta}g_{\mu\nu}(x)\dot x^{\mu}\dot x^{\nu}+\frac{3}{4}\beta\right)}
\langle W_{x(\tau)}\rangle_{\rm CS}}\,.
\end{equation}
Therefore, the free energy $F_{\rm sc}=-\log Z_{\rm sc}$ at large $N$ 
takes the form (\ref{Fsc-lN}), with $B_{\rm sc}(\lambda)$ 
given by 
\begin{equation}
B_{\rm sc}(\lambda)=-\int_0^{\infty}\frac{d\beta}{\beta}
\int_
{x^{\mu}(0)=x^{\mu}(1)}\!\!\!\!\!\!\!\!\!\!\!\!\!\!\!\!
{\cal D}x\,e^{-\int_0^1 d\tau \left(\frac{1}{4\beta}g_{\mu\nu}(x)\dot x^{\mu}\dot x^{\nu}+\frac{3}{4}\beta\right)}\,
\langle W_{x(\tau)}\rangle_{\rm CS}\,,
\label{BS-worldline}
\end{equation}
where the Wilson loop expectation value is computed in pure CS theory in the planar limit. 
Therefore, we find that we can obtain $B_{sc}(\lambda)$ by computing the expectation 
value of a Wilson loop supported on the loop $x(\tau)$ in pure CS theory, and then averaging 
over all closed loops using the worldline action in (\ref{BS-worldline}). 
Note that this representation of the planar free energy is valid in principle for any manifold, 
not just $S^3$ (more generally, one can add a curvature coupling $\beta \xi R$ and mass term $\beta m^2$ to 
the quantum mechanics action). A similar worldline approach was used in \cite{Armoni:2004ub,Armoni:2009jn} in the context of 
QCD in four dimensions. 

We now attempt to exactly evaluate the worldline path-integral in (\ref{BS-worldline}) in the case of $S^3$. 
As in the free case reviewed above, the calculation should take a form of a sum over the classical saddle points, 
and a path-integral around each classical solution. Namely, we can write
\begin{equation}
x(\tau) = x_n(\tau)+\sqrt{\beta}\xi(\tau)
\end{equation}
where $x_n(\tau)$ is a geodesic corresponding to a multiply wound great circle, path-integrate over $\xi(\tau)$, 
and sum over all $n$. For each loop, we have then to compute the Wilson loop expectation value 
$\langle W_{x_n(\tau)+\sqrt{\beta}\xi(\tau)}\rangle_{\rm CS}$. But since the pure CS theory is topological, this expectation 
value should not depend on the continuous parameter $\beta$ that controls the perturbative expansion. Therefore, 
we expect, at least naively, that we may replace 
\begin{equation}
\langle W_{x_n(\tau)+\sqrt{\beta}\xi(\tau)}\rangle_{\rm CS} = \langle W_{x_n(\tau)}\rangle_{\rm CS} \,.
\label{Wsimple}
\end{equation}
In other words, generically, we expect that the fluctuations $\xi(\tau)$ would produce some kind of ``wavy" deformation 
of the great circle, and the Wilson loop expectation value on such a wavy line in pure CS theory 
should be the same as the one of the circular ``unknot" $x_n(\tau)$. 
One may worry about whether non-generic deformations $\xi(\tau)$ could
introduce self-intersections, or make the loop knotted. In this case, the expectation 
value would depend on the topology of the knot. 
It would be important to study this more carefully. Here we will assume that (\ref{Wsimple}) holds, 
and proceed under this assumption. Then, the insertion of the Wilson 
loop in (\ref{BS-worldline}) does not affect the path-integral over quantum fluctuations around the $n$-wound 
circular geodesic, and just 
modifies the classical contribution of each geodesic. Denoting 
$\langle W_{x_n(\tau)}\rangle_{\rm CS} \equiv \langle W_n \rangle$, the result then should take the same form 
as in (\ref{logDetfree}) 
\begin{equation}
B_{\rm sc}(\lambda) = -2\pi^2\int_0^{\infty} \frac{d\beta}{\beta}\, \frac{e^{-\frac{3}{4}\beta}}{(4\pi\beta)^{3/2}}\left(e^{\beta}
+2\sum_{n=1}^{\infty} e^{-\frac{n^2\pi^2}{\beta}}e^{\beta}\left(1-\frac{2n^2\pi^2}{\beta}
\right)\langle W_n \rangle \right)\,.
\label{CSsc-beta}
\end{equation}
Integrating over $\beta$ (and dropping some purely imaginary term), we then get
\begin{equation}
B_{\rm sc}(\lambda) = -\sum_{n=1}^{\infty} (-1)^{n}\left(\frac{1}{4n}
-\frac{1}{2\pi^2 n^3}\right)\langle W_n \rangle\,.
\label{Bsc-CS}
\end{equation}

A similar calculation applies to the fermion theory. The generalization of the ${\cal N}=1$ SUSY quantum mechanics 
action (\ref{Ffree-fer}) in the presence of a background gauge fields yields the following representation 
for the one-loop determinant (for $m=0$)
\begin{equation}
\begin{aligned}
&-\frac{1}{2}\log\det\left (\slash \!\!\!\!\nabla^2_A\right)\\
&=\frac{1}{2}\int_0^{\infty}\frac{d\beta}{\beta}\int_{PBC}
{\cal D}x\, \int_{ABC} {\cal D}\psi {\rm tr}{\cal P}\,e^{-\int_0^1 d\tau \left(\frac{1}{4\beta}g_{\mu\nu}(x)\dot x^{\mu}\dot x^{\nu}+\frac{1}{2}\psi^{\mu}D_{\tau}\psi^{\nu}g_{\mu\nu}(x)-i\dot x^{\mu}A_{\mu}(x)
+i\beta \psi^{\mu}F_{\mu\nu}\psi^{\nu}\right)}\,.
\end{aligned}
\end{equation}
Inserting this into the partition function (\ref{Zbf}), and using large $N$ factorization as discussed above, we find 
$F_{\rm fer} = N^2 F_{CS}(\lambda)+N B_{\rm fer}(\lambda)+\ldots$ with
\begin{equation}
B_{\rm fer}(\lambda)=\frac{1}{2}
\int_0^{\infty}\frac{d\beta}{\beta}\int
{\cal D}x\, {\cal D}\psi \,e^{-\int_0^1 d\tau \left(\frac{1}{4\beta}g_{\mu\nu}(x)\dot x^{\mu}\dot x^{\nu}
+\frac{1}{2}\psi^{\mu}D_{\tau}\psi^{\nu}g_{\mu\nu}(x)\right)}\langle\frac{1}{N}{\rm tr}{\cal P} 
e^{i\oint (\dot x^{\mu}A_{\mu}-\beta \psi^{\mu}F_{\mu\nu}\psi^{\nu})}  \rangle_{\rm CS}\,.
\label{Bf-wl}
\end{equation}
Note that in this case we get a ``super"-Wilson loop that depends both on the closed path $x^{\mu}(\tau)$ and 
the antiperiodic fermionic variable $\psi^{\mu}(\tau)$. However, expanding this operator in powers of $\psi^{\mu}$ 
inserts 
factors of the field strength into the ordinary Wilson loop operator, and in turn such insertions are equivalent to 
adding small deformations of the contour. Due to the topological nature of CS theory, we then expect that (at least 
for smooth loops without self-intersections) the super-Wilson loop operator in (\ref{Bf-wl}) is in fact independent 
of $\psi^{\mu}$, and can be replaced by the ordinary Wilson loop operator. Assuming the arguments above 
leading to (\ref{Wsimple}) apply, we can then simply insert the expectation value $\langle W_n\rangle$ of the 
$n$-wound circular loop into the free field result (\ref{Kf-coinc}), and end up with 
\begin{equation}
\begin{aligned}
B_{\rm fer}(\lambda)
=\frac{1}{2}\int_0^{\infty}\frac{d\beta}{\beta} \frac{2\pi^2}{(4\pi \beta)^{\frac{3}{2}}}
\left(2-\beta+2\sum_{n=1}^{\infty} (-1)^n e^{-\frac{n^2\pi^2}{\beta}}\left(2-\beta-\frac{4n^2\pi^2}{\beta}
\right)\langle W_n\rangle \right)
\end{aligned}
\label{Bfer-wl}
\end{equation}
which yields 
\begin{equation}
B_{\rm fer}(\lambda) = -\sum_{n=1}^{\infty} (-1)^{n}\left(\frac{1}{4n}
+\frac{1}{2\pi^2 n^3}\right)\langle W_n \rangle\,.
\label{Bfer-CS}
\end{equation}

As a check of (\ref{Bsc-CS}) and (\ref{Bfer-CS}), we can compare to the expected supersymmetric result using the large $N$ 
relation (\ref{Bsu-to-Bsf}). This gives 
\begin{equation}
B_{\rm susy}(\lambda)=B_{\rm sc}(\lambda)+B_{\rm fer}(\lambda) 
= -\sum_{n=1}^{\infty} \frac{(-1)^{n}}{2n} \langle W_n \rangle\,,
\label{Bsusy-sum}
\end{equation}
which is indeed in precise agreement with the localization prediction (\ref{Bsusy-W}). 

To evaluate the sums explicitly, we need the expression for the expectation value of a multiply wound unknot. 
It is well-known that Wilson loop expectation values in CS theory depend on a choice of 
framing \cite{Witten:1988hf, Guadagnini:1989am, Labastida:2000yw, Marino:2001re,Brini:2011wi}. This is
essentially a choice of regularization that allows to define the ``self-linking" integrals of loops in 
a way consistent with topological invariance, and is labelled by an 
integer $f\in \mathbb{Z}$ (which represents the linking number of the loop $C$ with its slightly displaced framing loop $C_f$). 
With the simplest choice of framing (the more general case will be discussed below), 
corresponding to $f=0$, the expectation value of the $n$-wound unknot at large $N$ takes the form \cite{Witten:1988hf, Ooguri:1999bv}
\begin{equation}
\langle W_n^{f=0} \rangle = \frac{\sin(n\pi\lambda)}{n\pi\lambda} \,. 
\label{W0}
\end{equation}
Let us first consider the supersymmetric case. Inserting (\ref{W0}) into (\ref{Bsusy-sum}) gives
\begin{equation}
B_{\rm susy}(\lambda) = -\sum_{n=1}^{\infty} \frac{(-1)^{n}}{2n}\frac{\sin(n\pi\lambda)}{n\pi\lambda} = 
\frac{i}{4\pi\lambda}\left({\rm Li}_2(-e^{i\pi\lambda})-{\rm Li}_2(-e^{-i\pi\lambda})\right)
\label{BsusyW}
\end{equation}
which agrees with the result (\ref{Bsu}) obtained from the large $N$ limit of the matrix model. 

In the non-supersymmetric scalar and fermion theory, using the $f=0$ result (\ref{W0}) into 
(\ref{Bsc-CS}) and (\ref{Bfer-CS}), one gets 
\begin{eqnarray}
B_{\rm sc}^{f=0}(\lambda) &=& -\sum_{n=1}^{\infty} (-1)^{n}\left(\frac{1}{4n}
-\frac{1}{2\pi^2 n^3}\right)\frac{\sin(n\pi\lambda)}{n\pi\lambda} \cr 
&=& \frac{i}{8\pi\lambda}\left({\rm Li}_2(-e^{i\pi\lambda})-{\rm Li}_2(-e^{-i\pi\lambda})\right)
-\frac{i}{4\pi^3\lambda}\left({\rm Li}_4(-e^{i\pi\lambda})-{\rm Li}_4(-e^{-i\pi\lambda})\right)\,,
\label{B0sc}
\end{eqnarray}
and
\begin{eqnarray}
B_{\rm fer}^{f=0}(\lambda) &=& -\sum_{n=1}^{\infty} (-1)^{n}\left(\frac{1}{4n}
+\frac{1}{2\pi^2 n^3}\right)\frac{\sin(n\pi\lambda)}{n\pi\lambda} \cr 
&=& \frac{i}{8\pi\lambda}\left({\rm Li}_2(-e^{i\pi\lambda})-{\rm Li}_2(-e^{-i\pi\lambda})\right)
+\frac{i}{4\pi^3\lambda}\left({\rm Li}_4(-e^{i\pi\lambda})-{\rm Li}_4(-e^{-i\pi\lambda})\right)\,.
\label{B0fer}
\end{eqnarray}
These have the following behavior at small $\lambda$ and as $\lambda \rightarrow 1$
\begin{equation}
\begin{aligned}
&B_{sc}^{f=0}(\lambda)= \frac{\log 2 }{4}-\frac{3\zeta(3)}{8\pi^2} -\left(\frac{\pi^2}{96}-\frac{\log 2}{12}\right)\lambda^2-
\left(\frac{\pi^4}{3840}+\frac{\pi^2}{960}\right)\lambda^4+\ldots\\
&B_{sc}^{f=0}(\lambda)=-\frac{1}{4}(1-\lambda)\log (1-\lambda) +\ldots \qquad \lambda \rightarrow 1
\end{aligned}
\end{equation}
and 
\begin{equation}
\begin{aligned}
&B_{\rm fer}^{f=0}(\lambda)= \frac{\log 2}{4}+\frac{3\zeta(3)}{8\pi^2} -\left(\frac{\pi^2}{96}+\frac{\log 2}{12}\right)\lambda^2-
\left(\frac{\pi^4}{3840}-\frac{\pi^2}{960}\right)\lambda^4+\ldots\\
&B_{\rm fer}^{f=0}(\lambda)=-\frac{1}{4}(1-\lambda)\log (1-\lambda) +\ldots \qquad \lambda \rightarrow 1
\end{aligned}
\end{equation}
Note that the $\lambda\rightarrow 1$ behavior {\it does not} agree with the one predicted by the duality, eq. (\ref{lam1}). Indeed, one 
can verify that (\ref{B0sc}) and (\ref{B0fer}) do not satisfy the duality relation (\ref{bf-duality}).  

A potential resolution to this disagreement can be obtained by studying the dependence of the result (\ref{Bsc-CS}) and 
(\ref{Bfer-CS}) on the choice of framing parameter $f\in \mathbb{Z}$. The multiply wound unknot expectation value with framing $f$ 
is given for finite $N$ in \cite{Brini:2011wi} (see also \cite{Bianchi:2016gpg}). In the large $N$ 't Hooft limit, it reduces to
\begin{eqnarray}
\langle W^f_{n}\rangle 
&=& 
\frac{e^{i\pi  (f-1)n \lambda }}{2 \pi i \lambda } 
\sum_{l=0}^n \frac{ e^{2 \pi i l\lambda} (-1)^{l+n} (f n+l-1)!}{l! (n-l)! ((f-1) n+l)!} \cr
&=&\frac{e^{i\pi  (f-1)n \lambda }}{2 \pi i \lambda } 
\frac{(-1)^n (f n-1)!}{n! (f n-n)!}\, _2F_1\left(-n,f n,(f-1)n+1,e^{2 i \pi \lambda}\right)\,.
\label{Wnf}
\end{eqnarray}
This result may also be written in terms of the Jacobi polynomials $P_n^{(\alpha,\beta)}(x)$ as 
\begin{equation}
\langle W^f_{n}\rangle = \frac{e^{i\pi  (f-1)n \lambda }}{2 \pi i \lambda f n } 
(-1)^n P_n^{(f n-n,-1)}(1-2 e^{2 i \pi  \lambda }).
\end{equation}
One can check that for $f=0$, this reduces to (\ref{W0}). Another notable value is $f=1$: this is the choice of framing which is 
automatically picked by the localization calculation, see e.g. \cite{Kapustin:2009kz} and the Appendix. Note that 
for $n=1$ (\ref{Wnf}) gives $\langle W_{n=1}^f\rangle = e^{i\pi f \lambda}\frac{\sin(\pi\lambda)}{\pi\lambda}$, but for general 
$n$ the framing dependence is not simply through an overal phase, see for instance eq. (\ref{Wnloc}) for the $f=1$ case.\footnote{One may write a multiply wound Wilson loop as a sum 
of Wilson loops in hook representations \cite{Marino:2001re}. 
For each irreducible representation, the framing dependence is through an overall phase only, 
but the phase factor depends on the representation. The somewhat complicated framing dependence of the 
multiply wound loop (\ref{Wnf}) reflects this fact.}  

Using (\ref{Wnf}), one can verify that, remarkably, the term in the sum (\ref{Bsc-CS})-(\ref{Bfer-CS}) involving 
$(-1)^n/n$ is independent of the framing
\begin{equation}
-\sum_{n=1}^{\infty} \frac{(-1)^{n}}{4n} \langle W_n^{f} \rangle = 
-\sum_{n=1}^{\infty} \frac{(-1)^{n}}{4n} \langle W_n^{f=0} \rangle \,.
\label{n1-f}
\end{equation}
More precisely, the above equality holds up to a purely imaginary piece linear in $f$. Here and below, we will 
consistently drop imaginary terms and focus on the real part of the free energy (this can be done for instance 
by averaging the $f$ and $-f$ framings). Note that (\ref{n1-f}) implies in particular that 
(the real part of) the SUSY result (\ref{BsusyW}) is independent of $f$. 

On the other hand, the sum involving $(-1)^n/n^3$ turns out to depend on $f$, but in a simple quadratic 
way
\begin{equation}
-\sum_{n=1}^{\infty} \frac{(-1)^{n}}{2\pi^2 n^3} 
(\langle W_n^{f} \rangle - \langle W_n^{f=0} \rangle)=  f^2 \Delta B(\lambda)
\end{equation} 
with 
\begin{equation}
\Delta B(\lambda) = 
-\frac{1}{24} \lambda ^2 (1+\log (4))
+\frac{\pi ^2 \lambda ^4}{192}
+\frac{\pi ^4 \lambda ^6}{6912}
+\frac{67 \pi ^6 \lambda ^8}{8709120}+\frac{41 \pi^8 \lambda^{10}}{82944000}+\ldots 
\label{dBser}
\end{equation}
and it is straightforward to compute higher orders in this expansion. Note that this result implies that the 
highest transcendentality terms of the small $\lambda$ expansion of (\ref{Bsc-CS}),(\ref{Bfer-CS}) are not affected 
by the framing parameter and are controlled by the SUSY result. 

Comparing with the small $\lambda$ expansion of (\ref{gsol}) and (\ref{B0sc}),(\ref{B0fer}), 
we find that the expansion (\ref{dBser}) of $\Delta B(\lambda)$ in fact corresponds to the closed form expression
\begin{equation}
\Delta B(\lambda) = \frac{g(\lambda)}{\lambda}
+\frac{i}{4 \pi ^3 \lambda}
\left(\text{Li}_4(-e^{i \pi  \lambda })-\text{Li}_4(-e^{-i \pi  \lambda })\right)
\end{equation}
with $g(\lambda)$ given in (\ref{gsol}). It then follows that we can obtain a result consistent with the duality relation (\ref{bf-duality}) by adding 
a ``correcting" term to the $f=0$ results (\ref{B0sc}),(\ref{B0fer}) as
\begin{eqnarray}
B_{\rm sc}(\lambda) &=& B_{\rm sc}^{f=0}(\lambda)+\Delta B(\lambda) = 
 -\sum_{n=1}^{\infty} (-1)^{n}\left(\frac{1}{4n}
-\frac{1}{2\pi^2 n^3}\right)
(2\langle W_n^{f=0} \rangle-\langle W_n^{f=1} \rangle)\cr 
 &=& \frac{1}{2}B_{\rm susy}(\lambda)+\frac{g(\lambda)}{\lambda}
\label{Bsc-presc}
\end{eqnarray}
and 
\begin{eqnarray}
B_{\rm fer}(\lambda) &=& B_{\rm fer}^{f=0}(\lambda)-\Delta B(\lambda) = 
 -\sum_{n=1}^{\infty} (-1)^{n}\left(\frac{1}{4n}
+\frac{1}{2\pi^2 n^3}\right)
(2\langle W_n^{f=0} \rangle-\langle W_n^{f=1} \rangle)\cr 
 &=& \frac{1}{2}B_{\rm susy}(\lambda)-\frac{g(\lambda)}{\lambda}\,.
\label{Bfer-presc}
\end{eqnarray}
These results coincide with the solution (\ref{Bsc-sol}), (\ref{Bfer-sol}) constructed in Section \ref{solmap}, and hence the corresponding 
free energies satisfy the duality (\ref{bf-duality}). While it seems suggestive that a relatively simple 
combinations of $f=0$ and $f=1$ framed Wilson loops yields a result consistent with duality, the origin of 
such prescription is unclear. One possibility is that the specific combination of Wilson loop expectation 
values appearing in (\ref{Bsc-presc}),(\ref{Bfer-presc}) effectively implements 
a refinement of the arguments around eq. (\ref{Wsimple}). 
A direct perturbative calculation of the free energy on $S^3$ would be very useful to further clarify these issues.

\subsection{Small mass expansion}
\label{massive-sec}

It is straightforward to generalize the calculations of the previous section by adding a mass term. This can be done by including a factor 
$e^{-\beta m^2}$ in the proper time integrals (\ref{CSsc-beta}) and (\ref{Bfer-wl}). In the scalar case, after integrating over $\beta$ we get
\begin{equation}
B_{\rm sc}(\lambda,m^2)  =-\frac{\pi}{3}(m^2-\frac{1}{4})^{\frac{3}{2}}
-\sum_{n=1}^{\infty}\,e^{-2\pi n \sqrt{m^2-\frac{1}{4}}}\frac{2n^2\pi^2(\frac{1}{4}-m^2)
-1-2\pi n \sqrt{m^2-\frac{1}{4}}}{2\pi^2 n^3}\,\langle W_n\rangle \,.
\label{Bsc-m}
\end{equation}

For non-zero mass, the large $N$ sphere free-energies of ``regular" and ``critical" scalar are different. Let us first consider the regular 
scalar theory (the critical case is discussed in the next section). In this theory, we can also add a $(\bar\phi\phi)^3$ interaction and study the 
dependence on the corresponding coupling constant $\lambda_6$. Introducing a pair of auxiliary fields, we can replace
\begin{equation}
S_{\lambda_6}=\int d^3 x \sqrt{g} \frac{\lambda_6}{3!N^2}(\bar\phi \phi)^3 \rightarrow
\int d^d x \sqrt{g}\left[\sigma(\bar\phi\phi-\rho)+\frac{\lambda_6}{3!N^2}\rho^3\right]
\label{rhosig}
\end{equation} 
The path-integral over $\sigma$ sets $\bar\phi\phi = \rho$, yielding back the sextic interaction. In the large $N$ limit, we 
can compute the free energy as a function of $\rho$ and $\sigma$, and look for 
saddle point solutions with constant $\rho$ and $\sigma$. Therefore, we may directly eliminate $\sigma$ using the classical 
equations of motion, and work with
\begin{equation}
S_{\lambda_6}=\int d^3 x \sqrt{g}\left[\frac{\lambda_6}{2}\rho^2 \phi^{\dagger}\phi-\frac{N\lambda_6}{3}\rho^3\right]
\end{equation}
where we have rescaled $\rho$ by a factor of $N$ for convenience. The order $N$ term in the free energy of the regular scalar 
theory is then given by 
\begin{equation}
B_{\rm sc}(\lambda,\lambda_6,m^2)=B_{\rm sc}(\lambda,m^2+\frac{\lambda_6}{2}\rho_*^2)-{\rm vol}(S^3) \frac{\lambda_6}{3}\rho_*^3
\label{Fg6}
\end{equation}
where $B_{\rm sc}(\lambda,m^2)$ is given in (\ref{Bsc-m}), and $\rho_*$ is the solution to the saddle point equation 
for $\rho$: 
\begin{equation}
\frac{d}{d\rho}\left[B_{\rm sc}(\lambda,m^2+\frac{\lambda_6}{2}\rho^2)-{\rm vol}(S^3) \frac{\lambda_6}{3}\rho^3\right]=0\,.
\label{sig-saddle}
\end{equation}
It is interesting to derive the small mass expansion of the free energy, since its coefficients are related to the integrated correlation 
functions of the $j_0=\bar\phi\phi$ operator. This can be done by solving (\ref{sig-saddle}) in powers of $m^2$, 
and plugging back into (\ref{Fg6}).  It will be convenient to introduce the notation
\begin{equation}
{\cal S}_k \equiv  \sum_{n=1}^{\infty} (-1)^n n^k \langle W_n \rangle 
\end{equation}
for the Wilson loop sums which arise when expanding (\ref{Bsc-m}) at small $m$. One can show that ${\cal S}_0 = -1/2$ 
and ${\cal S}_{2k}=0$ for $k>0$ (we have checked that this holds for any choice of framing). Solving (\ref{sig-saddle}) 
we find $\rho_* = \frac{m^2}{2} {\cal S}_1 -\frac{m^4}{16}(8 {\cal S}_1-\lambda_6 {\cal S}_1^3)+\ldots$, and plugging this into (\ref{Fg6}) 
yields 
\begin{equation}
B_{\rm sc}(\lambda,\lambda_6,m^2) = B_{\rm sc}(\lambda) +\frac{\pi^2}{2}{\cal S}_1 m^4 
+ \frac{\pi^2}{24}\left(8 {\cal S}_1-\lambda_6 {\cal S}_1^3\right) m^6+O(m^8) 
\label{Bf-regsc}
\end{equation}
where $B_{\rm sc}(\lambda)$ is the free energy in the conformal case discussed in the previous section. Note that the term of order $m^2$ drops 
out, consistently with conformal invariance (this follows from ${\cal S}_0 = -1/2$ and holds for any choice of framing). 

Let us specialize to the simplest choice of framing of the Wilson loop, $f=0$. Then we find
\begin{equation}
{\cal S}_1 = \sum_{n=1}^{\infty} (-1)^n n \frac{\sin(n\pi\lambda)}{n\pi\lambda} =  -\frac{\tan(\frac{\pi\lambda}{2})}{2\pi\lambda}\,.
\label{S1-sum}
\end{equation}
From the term of order $m^4$ in (\ref{Bf-regsc}), and using the 2-point integral in (\ref{I2I3}), we can then read off the 
2-point function normalization to be 
\begin{equation}
\langle j_0 j_0\rangle_{\rm CS-sc} = \frac{2\tan(\frac{\pi\lambda}{2})}{\pi\lambda} 
\langle j_0 j_0\rangle_{{\rm free~sc}}
\label{J02pt}
\end{equation}
where the free correlator refers to the theory of $N$ complex scalar fields, with $j_0=\bar\phi \phi$.
This result precisely agrees with the explicit calculation in light-cone gauge done in \cite{Aharony:2012nh}. 
Interestingly, it arises here in a gauge independent way as a sum over Wilson loops (\ref{S1-sum}). 

From the term 
of order $m^6$ and the 3-point integral in (\ref{I2I3}), we read off the 3-point function normalization 
\begin{equation}
\langle j_0 j_0j_0\rangle_{\rm CS-sc}= \left[\frac{2\tan(\frac{\pi\lambda}{2})}{\pi\lambda} 
-\frac{\tan^3(\frac{\pi\lambda}{2})}{16\pi^3\lambda^3}\lambda_6\right]\langle  j_0 j_0j_0\rangle_{{\rm free~sc}}
\label{JJJg6}
\end{equation}
This has the same structure of the result obtained in \cite{Maldacena:2012sf} from weakly broken higher-spin symmetry,
using $\tilde\lambda = \tan(\frac{\pi\lambda}{2})$ \cite{Aharony:2012nh}, and provided one identifies the parameter 
$a_3$ defined in \cite{Maldacena:2012sf} to be $a_3 \propto \lambda_6$. However, the explicit calculation 
in light-cone gauge performed in \cite{Aharony:2012nh} gave $a_3 \propto \lambda_6-24\pi^2 \lambda^2$. In other 
words, we find that our result (\ref{JJJg6}) would agree precisely with the one in \cite{Aharony:2012nh} 
only if we redefine $\lambda_6 \rightarrow \lambda_6-24\pi^2\lambda^2$, or equivalently 
add to (\ref{rhosig}) a term $-\frac{4\lambda^2}{N^2}\rho^3$. One may view the ``regular" scalar theory 
defined by (\ref{rhosig}) as a Legendre transform of the critical scalar theory (see the next section), 
where, before making it dynamical, $\rho$ plays the role of source for the $\sigma$ correlators. Then modifying 
the coefficient of the $\rho^3$ term is equivalent to modifying the contact term in the $\sigma$ 3-point function, 
which is scheme dependent. 

Let us now move to the (regular) fermion theory. Including the $e^{-\beta m^2}$ factor in (\ref{Bfer-wl}) and 
integrating over $\beta$, we get 
\begin{equation}
B_{\rm fer}(\lambda,m)= 
\frac{\pi}{12} m(4m^2+3)
-\sum_{n=1}^{\infty} (-1)^n e^{-2n\pi m }\frac{n^2\pi^2(4m^2+1)+4n\pi m+2}{4\pi^2 n^3}\langle W_n\rangle\,. 
\end{equation}
The small mass expansion reads 
\begin{equation}
B_{\rm fer}(\lambda,m) = B_{\rm fer}(\lambda) -\frac{\pi^2}{2}{\cal S}_1 m^2 
-\frac{\pi^2}{6}(6{\cal S}_1+\pi^2 {\cal S}_3)m^4+\ldots 
\label{Bf-m}
\end{equation}
and we see that the term of order $m$ drops out, as expected from conformal invariance. 
Using the $f=0$ result (\ref{S1-sum}), and (\ref{I2I3}), we can read off from the $m^2$ term 
the 2-point function normalization of the 
$\tilde{j}_0 = \bar\psi \psi$ operator 
\begin{equation}
\langle \tilde{j}_0 \tilde{j}_0 \rangle_{\rm CS-fer} 
= \frac{2\tan(\frac{\pi\lambda}{2})}{\pi\lambda}
\langle \tilde{j}_0 \tilde{j}_0 \rangle_{\rm free~fer}
\end{equation}
which is again in agreement with the direct light-cone gauge calculation in \cite{GurAri:2012is}. Note that 
the $m^3$ term is absent in (\ref{Bf-m}). The 3-point function of $\bar\psi \psi$ is expected 
to be a pure contact term \cite{GurAri:2012is}, which 
is scheme dependent. The calculation above apparently picks a scheme where it vanishes. One 
may adjust such contact term by adding by hand a $m^3$ term to the free energy (viewing $m$ as the constant 
part of the source that couples to the operator $\bar\psi\psi$). 

\subsubsection{Critical theories}
\label{critTh}

The critical scalar theory coupled to Chern-Simons is obtained by adding 
the $\frac{\lambda_4}{4N} (\bar\phi \phi)^2$ term and flowing 
to the IR. In the large $N$ limit, we can describe this theory by introducing a Hubbard-Stratonovich 
auxiliary field
\begin{equation}
\frac{\lambda_4}{4N} (\bar\phi\phi)^2 \rightarrow \sigma (\bar\phi \phi)-\frac{N}{\lambda_4}\sigma^2
\end{equation}
In the IR limit, the quadratic term in $\sigma$ can be dropped, and one can work 
with the action 
\begin{equation}
S_{\rm crit} =S_{CS_{N,k}}
+\int d^3x \left( D_{\mu}\bar{\phi} D^{\mu}\phi+\sigma \bar\phi \phi \right)\,,
\label{crit-bos}
\end{equation}
At large $N$, we can perform the $\phi$ path-integral assuming constant $\sigma$, and extremize 
in $\sigma$ at the end. Using the result for the massive scalar in the previous section, we have
\begin{equation}
B_{\rm crit.~sc}(\lambda)=B_{\rm sc}(\lambda, m^2=\sigma)|_{\sigma=\sigma^*}
\end{equation}
where $\sigma^*$ is the value of $\sigma$ extremizing $B_{\rm sc}(\lambda, m^2=\sigma)$. 
But since the derivative of $B_{\rm sc}(\lambda, m^2)$ with 
respect to $m^2$ vanishes at $m=0$ (because one-point functions are zero in the conformal theory on $S^3$), 
then $\sigma^*=0$ and the critical scalar result coincides with the massless scalar one at large $N$, as 
also explained in Section \ref{BF-dual-sec} above 
\begin{equation}
B_{\rm crit.~sc}(\lambda)=B_{\rm sc}(\lambda)\,.
\end{equation}
When we add a mass deformation, this is no longer true. Let us therefore consider the mass deformed critical 
theory defined by 
\begin{equation}
S_{\rm crit}(m) =S_{CS_{N,k}}
+\int d^3x \left( D_{\mu}\bar{\phi} D^{\mu}\phi+\sigma \bar\phi \phi +N m \sigma \right)\,,
\label{crit-bos-m}
\end{equation}
At the level of the sphere free energy at large $N$, we then find 
\begin{equation}
B_{\rm crit.~sc}(\lambda,m) = \left[B_{\rm sc}(\lambda, \sigma )+{\rm vol}(S^3)m \sigma\right]_{\sigma=\sigma_*}
\label{Brit-m}
\end{equation}
where $\sigma_*$ is determined by extremizing 
\begin{equation}
\frac{d}{d\sigma} B_{sc}(\lambda, \sigma)= -m\,.
\end{equation}
Solving this equation perturbatively at small $m$ 
and plugging back in (\ref{Brit-m}), one finds 
\begin{equation}
B_{\rm crit.~sc}(\lambda,m) 
= B_{\rm sc}(\lambda) -\frac{2\pi^2}{{\cal S}_1}m^2 +\frac{8\pi^2}{3{\cal S}_1^2} m^3
-\frac{8\pi^2}{3{\cal S}_1^4}(6{\cal S}_1+\pi^2 {\cal S}_3)m^4+\ldots  
\label{crit-mass}
\end{equation}
Using the 2-point integral in (\ref{I2I3}) with $\Delta=2$, and the $f=0$ result (\ref{S1-sum}), 
the order $m^2$ term in the above expansion  
implies that the flat space 2-point function of $\sigma$ is 
\begin{equation}
\langle \sigma(x) \sigma (0) \rangle_{\rm CS-crit.sc.} = 
\frac{4\lambda\cot(\frac{\pi \lambda}{2})}{N \pi}\, \frac{1}{x^4}
\end{equation}
This agrees with the result in \cite{Aharony:2012nh} (it essentially follows from the regular theory result (\ref{J02pt}) 
by the Legendre transform). The 3-point function in the large $N$ critical scalar theory is 
expected to be a pure contact term \cite{Aharony:2012nh} (at $\lambda=0$, 
this is well-known \cite{Petkou:1994ad}).  
The non-zero coefficient of the term of order $m^3$ in (\ref{crit-mass}) 
should come from integration over the sphere of this contact term. Such contact terms are expected to be scheme dependent 
(from the free energy point of view, they can be changed by adding a term proportional to $m^3$). 

Finally, let us discuss the critical fermionic theory. This can be obtained from the ``regular" fermion theory by 
a Legendre transform
\begin{equation}
    S_{\rm crit-fer}=S_{CS_{N,k}}+\int d^3x \left(\bar{\psi} \slashed{D} \psi+\sigma \bar\psi\psi +N m^2 \sigma 
+\frac{N \lambda_6^{f}}{3!}\sigma^3\right)\,,
\label{Scritfer}
\end{equation}
where we have included an arbitrary mass parameter and sextic coupling ($\sigma$ plays the role of $\bar\psi\psi$ at 
the UV fixed point). The order $N$ term of the free energy is obtained from 
\begin{equation}
B_{\rm crit-fer}(\lambda,\lambda_6^f,m^2) 
= \left[B_{\rm fer}(\lambda,m=\sigma)+{\rm vol}(S^3) (m^2 \sigma + \frac{\lambda_6^f}{6}\sigma^3)\right]_{\sigma=\sigma_*}
\end{equation} 
where $\sigma_*$ is obtained by extremizing in $\sigma$, and $B_{\rm fer}(\lambda,m)$ is given in (\ref{Bf-m}). Solving 
for $\sigma_*$ in the small mass expansion, we find 
\begin{equation}
B_{\rm crit-fer}(\lambda,\lambda_6^f,m^2)  
= B_{\rm fer}(\lambda)+\frac{2\pi^2}{{\cal S}_1}m^4+  \frac{8\pi^2\lambda_6^f}{3 {\cal S}_1^3}m^6+\ldots 
\end{equation}
and using (\ref{S1-sum}) and (\ref{I2I3}) this means
\begin{equation}
\begin{aligned}
&\langle \sigma(x)\sigma(0)\rangle_{\rm CS-crit.fer.} = 
\frac{2\lambda \cot(\frac{\pi\lambda}{2})}{N\pi} \frac{1}{x^2}\\
&\langle \sigma(x_1)\sigma(x_2)\sigma(x_3)\rangle_{\rm CS-crit.fer.} = -8\lambda_6^f \lambda^3 \cot^3(\frac{\pi\lambda}{2})
\frac{1}{|x_{12}| |x_{23}| |x_{31}|}
\end{aligned}
\end{equation}
The 2-point function agrees with the expected result \cite{GurAri:2012is}. The $\lambda_6^f$ dependence of the 
3-point function also precisely agrees with the one derived in \cite{GurAri:2012is}, however the result 
given there also includes an additional $\lambda_6^f$-independent term that originates, via the Legendre transform,
from the 3-point function of $\bar\psi\psi$ in the regular fermion theory, which is a pure contact term and hence 
expected to be scheme dependent. Similarly to the case of the regular scalar theory discussed above, in 
the critical fermion theory (\ref{Scritfer}) one may adjust such contribution 
by adding an extra term proportional to $\sigma^3$, or equivalently redefining $\lambda_6^f$ by a finite $\lambda$-dependent 
shift.

\section*{Acknowledgments}
I would like to thank O. Aharony, I. Klebanov, S. Minwalla, B. Safdi, G. Tarnopolsky, I. Yaakov and R. Yacoby for many valuable discussions. 
I am especially grateful to O. Aharony, I. Klebanov and S. Minwalla for useful comments on a draft of this paper. 
This work is supported in part by the US NSF under Grant No.~PHY-1620542.

\begin{appendices}

\section{Free energy of pure Chern-Simons theory at large $N$}

 First, let us recall that, for $U(N)$ Chern-Simons gauge theory with level $k$ \cite{Witten:1988hf, Marino:2005sj, Marino:2011nm}
\begin{equation}
  {\cal F}_\text{CS}(N,k)= -\log Z_{\text{CS}(N,k)} = \frac{N}{2} \log (k+N) - \sum_{j=1}^{N-1} (N-j) \log \left (2 \sin \frac{\pi j}{k+N} \right )
   \,.
\label{FCS}
\end{equation}
More precisely, this is the free energy of the $U(N)_{k,k+N}$ theory, i.e. with the level of the $U(1)$ factor taken to be $k+N$. 
One can check that it exhibits the level-rank duality
\begin{equation}
{\cal F}_\text{CS}(N, k)= {\cal F}_\text{CS}(k,N)
\ .
\end{equation}
The free energy for the $SU(N)_k$ theory 
can be obtained from the one in (\ref{FCS}) by subtracting the $U(1)$ factor $\frac{1}{2}\log((k+N)/N)$, see e.g. \cite{Marino:2005sj}. For 
the $U(N)_{k,k'}$ theory, with $k'$ being the level of the $U(1)$ factor, one has ${\cal F}_{U(N)_{k,k'}} = {\cal F}_{SU(N)_k}+1/2\log(k'/N)$. One 
can verify that ${\cal F}_{U(N)_{k,k'}}$ and ${\cal F}_{SU(N)_k}$ satisfy all the relevant level/rank dualities listed in \cite{Aharony:2015mjs,Hsin:2016blu}.

We are interested in the expansion of the free energy (\ref{FCS}) in the 't Hooft limit
\begin{equation}
N\rightarrow \infty\,,\quad k\rightarrow \infty\,,\qquad \lambda=\frac{N}{k+N}\,\,\mbox{fixed} \,.
\end{equation}
The free energy has the large $N$ expansion\footnote{To be more precise, there is also a term of order $\log N$ that 
essentially comes 
from the path integral measure, see \cite{Marino:2005sj}. We can formally incorporate it in $F^{(1)}_{\rm CS}(\lambda)$, which 
has the explicit expression 
$F^{(1)}_{\rm CS}(\lambda) = \frac{1}{24} \left(\log \left(1-e^{-2 i \pi  \lambda }\right)
+\log \left(1-e^{2 i \pi  \lambda }\right)\right)
+\frac{1}{12} \log \left(\frac{2 \pi  N}{\lambda }\right)-\zeta '(-1)$ \cite{Marino:2005sj, Hatsuda:2015owa}.}
\begin{equation}
{\cal F}_{\rm CS}(N,\lambda)=N^2 F_{\rm CS}(\lambda)+F^{(1)}_{\rm CS}(\lambda)
+\frac{1}{N^2}F^{(2)}_{\rm CS}(\lambda)+\ldots
\end{equation}
The leading term of order $N^2$ can be easily extracted using the fact that
\begin{equation}
\lim_{N\rightarrow \infty} \frac{1}{N}\sum_{j=1}^N f(\frac{j}{N}) = \int_0^1 dx f(x)\,.
\end{equation}
Then we have
\begin{equation}
\begin{aligned}
&\sum_{j=1}^{N-1}(N-j)\log(2\sin\frac{\pi j}{k+N})\\
&=N\sum_{j=1}^{N-1}(1-\frac{j}{N})\log(2\sin\frac{\pi j\lambda}{N})
\stackrel{N\rightarrow \infty}{\rightarrow}N^2\int_0^1 dx(1-x)\log(2\sin(\pi \lambda x))\,.
\end{aligned}
\end{equation}
So we find
\begin{equation}
F_{\rm CS}(\lambda)=-\int_0^1 dx(1-x)\log(2\sin(\pi \lambda x))\,.
\label{A-int}
\end{equation}
This result agrees with the one obtained in \cite{Camperi:1990dk}. We can evaluate the integral for instance by using 
\begin{equation}
\begin{aligned}
\log(2\sin(\pi \lambda x)) = -\sum_{\ell=1}^{\infty} \frac{\cos (2\pi \lambda \ell x)}{\ell}\,.
\end{aligned}
\end{equation}
Inserting this result into (\ref{A-int}) gives
\begin{equation}
\begin{aligned}
&F_{\rm CS}(\lambda)=\int_0^1 dx(1-x)\sum_{\ell=1}^{\infty} \frac{\cos (2\pi \lambda \ell x)}{\ell}=\sum_{\ell=1}^{\infty} \frac{\sin^2(\pi \lambda \ell)}{2\pi^2\lambda^2\ell^3}\\
&=\frac{1}{8\pi^2\lambda^2}\left(2\zeta(3)-\mbox{Li}_3(e^{2i\pi\lambda})-\mbox{Li}_3(e^{-2i\pi\lambda})\right) \,,
\label{A-final}
\end{aligned}
\end{equation}
which agrees with the result given in~\cite{Gopakumar:1998ki} (see also \cite{Ooguri:1999bv,Marino:2005sj}). 
It is straightforward to verify that $F_{\rm CS}(\lambda)$ satisfies the level-rank duality, which at 
large $N$ amounts to (\ref{CS-duality}). 
We also note that the small $\lambda$ expansion of~\eqref{A-final} reads
\begin{equation}
F_{\rm CS}(\lambda)=\frac{3}{4}-\frac{1}{2}\log(2\pi\lambda)+\frac{\pi^2\lambda^2}{72}+\frac{\pi^4\lambda^4}{5400}
+\frac{\pi ^6 \lambda^6}{158760}+\ldots \,.
\label{A-pert}
\end{equation}


\section{Multiply wound Wilson loop from localization}

The expectation value of the multiply wound circular loop in Chern-Simons theory at large $N$ 
can be obtained in the localization approach as the integral
\begin{equation}
\langle W_{n}\rangle = \int_{-a}^a du\rho_0(u) e^{n u}= \int_{-a}^a du\rho_0(u) \cosh(n u)\,,
\end{equation}
where $a= 2 \cosh^{-1} \exp(t/2)$ and the eigenvalue density is given in (\ref{density}), 
and recall that $t$ is related to the 't Hooft coupling by analytic continuation, $t=2\pi i \lambda$. Performing the integral by similar 
methods as in (\ref{bEval}) for real positive $t$, and then analytically continuing to imaginary $t$, one gets 
\begin{equation}
\begin{aligned}
 &\langle W_{n=1}\rangle =  e^{i\pi\lambda} \frac{\sin(\pi\lambda)}{\pi \lambda}\,,\qquad 
 \langle W_{n=2}\rangle = 3e^{2i\pi\lambda}\frac{\sin(2\pi\lambda)}{2\pi \lambda}
-2e^{i\pi\lambda} \frac{\sin(\pi\lambda)}{\pi \lambda}\,,\\
&\langle W_{n=3}\rangle =10 e^{3 i \pi  \lambda } \frac{\sin (3 \pi  \lambda )}{3 \pi  \lambda } 
-12 e^{2 i \pi  \lambda } \frac{\sin (2 \pi  \lambda )}{2\pi  \lambda }+ 3 e^{i \pi  \lambda } \frac{\sin (\pi  \lambda )}{\pi  \lambda }\,,\quad \ldots 
\label{Wnloc}
\end{aligned}
\end{equation}
One can verify that these results agree with the $f=1$ case of the general framing answer given in (\ref{Wnf}). The reason for the appearance of the $f=1$ 
choice of framing in the localization framework was explained in \cite{Kapustin:2009kz}.

\end{appendices}
\bibliographystyle{ssg}
\bibliography{CSBib}

\end{document}